\documentclass[twoside]{article}

\usepackage[preprint]{aistats2027}
\usepackage[round]{natbib}

\usepackage{stfloats}
\usepackage[utf8]{inputenc} 
\usepackage[T1]{fontenc}    
\usepackage{hyperref}       
\usepackage{url}            
\usepackage{booktabs}       
\usepackage{amsfonts}       
\usepackage{nicefrac}       
\usepackage{microtype}      
\usepackage{xcolor}         
\usepackage{amsmath}
\usepackage{amssymb}
\usepackage{mathtools}
\usepackage{amsthm}
\usepackage{accents}
\usepackage {tikz}
\usetikzlibrary {positioning}
\usetikzlibrary{trees}
\usepackage{microtype}
\usepackage{graphicx}
\usepackage{subcaption}
\usepackage{booktabs}
\usepackage{comment}
\input{mysymbol.sty}

\theoremstyle{plain}
\newtheorem{theorem}{Theorem}
\newtheorem{proposition}[theorem]{Proposition}
\newtheorem{lemma}[theorem]{Lemma}

\theoremstyle{definition}

\theoremstyle{remark}

\begin{document}

%

%

\twocolumn[

\aistatstitle{Spectral Convergence of Sparse Bipartite Graphs via the Line-Graphon}

\aistatsauthor{ Roxanne Holden \And Luana Ruiz}

\aistatsaddress{ Johns Hopkins University\\Baltimore, MD, USA} ]

\begin{abstract}
Pure point spectral convergence is generally unavailable for sparse graph sequences, where Laplacian eigenvalues may remain unstable under graph growth. This poses challenges for analyzing spectral operators underlying many graph-based learning methods, including graph neural networks. We identify a class of sparse bipartite graph sequences for which eigenvalue-level spectral convergence can be recovered. Focusing on graphs with strong degree heterogeneity, we exploit the convergence properties of their line-graph sequences. Using a spectral correspondence between bipartite graph Laplacians and line-graph adjacency operators, we transfer convergence guarantees from dense line-graphs back to the original sparse graphs. This framework yields eigenvalue-level spectral convergence for a class of deterministic sparse graph sequences. We further analyze superlinear preferential attachment trees, establishing convergence to an infinite-dimensional limit operator. Numerical experiments empirically support our spectral convergence results.


\end{abstract}

\section*{Introduction}
Spectral convergence provides a foundation for analyzing graph Laplacians and related operators in large-graph regimes. For dense graphs, suitably normalized graph operators converge to compact limit operators with pure-point spectra, enabling convergence of individual eigenvalues
\citep{borgs2008convergent, lovasz2012large, ruiz20-transf,belkin2003laplacian}. In contrast, for sparse graphs, such convergence generally fails. Sparse graphs may exhibit highly unstable eigenvalues under graph growth, and existing limit theories typically yield non-compact operators with continuous spectra, rather than pure point convergence of individual eigenvalues \citep{Salez2020SpectraSparse}. As a result, eigenvalue-by-eigenvalue comparisons across sparse graphs of different sizes are generally not well-posed without additional structural assumptions \citep{Salez2020SpectraSparse, lovasz2012large}.

This limitation is relevant in machine learning on graphs. Many graph-based learning methods, including graph neural networks (GNNs) \citep{kipf17-classifgcnn, defferrard17-cnngraphs, bruna14-deepspectralnetworks}, graph filters \citep{segarra17-linear,sandryhaila2013discrete}, and diffusion-based regularization schemes \citep{belkin2006manifold,calder2023rates,pang2017graph}, act as spectral operators of the graph Laplacian. Their consistency and transferability depend on the stability of the underlying Laplacian spectrum. Spectral instability under graph growth consequently complicates reasoning about transferability and robustness across graphs of different sizes \citep{gama19-stability,ruiz20-transf,zhang2025stability}. 


Yet, spectral instability does not affect all sparse graphs uniformly. We identify two structural conditions under which pure point spectral convergence can be recovered for sparse graphs: bipartiteness and high degree heterogeneity with a mild regularity condition on the degree profile \citep{kandanaarachchi2024graphons}. Sparse graphs often exhibit approximately bipartite structure \citep{cranston2020sparse}, and high degree heterogeneity holds for popular graph families in networks science such as star graphs and superlinear preferential attachment (SPA). 


Sparse graphs satisfying the square-degree property, which enforces strong degree heterogeneity, admit a dense representation via their line-graphs \citep{kandanaarachchi2024graphons}. In these representations, edges of the original graph become nodes and hubs induce dense local structure. Thus, the associated line-graphs admit graphon limits, yielding compact integral operators with pure point spectra and enabling eigenvalue-level spectral analysis \citep{borgs2008convergent}.

A second component of our approach is a direct spectral correspondence between the Laplacian of a bipartite graph and the adjacency operator of its line-graph. It can be shown that their nonzero spectra differ by a constant shift (Lemma~\ref{eig}), which allows spectral convergence results obtained at the line-graphon level to be transferred back to the original sparse bipartite graphs.

Building on the line-graph representation, we establish eigenvalue-level convergence guarantees for sparse bipartite graph sequences with mild degree profile regularity. Under minimal assumptions, we prove asymptotic convergence for Laplacian eigenvalues (Thm.~\ref{laplacesparse}). As a case study, we analyze SPA trees and show convergence of their Laplacian spectra to that of a compact limit operator (Thm.~\ref{SPAcor}). Numerical experiments on synthetic and real-world graphs illustrate these results and their implications for GNN transferability. 
\label{intro}

\section*{Preliminary Definitions}
{Graphs} are pairs \( G = (V, E) \) consisting of a node set \( V \) and an edge set \( E \subseteq V \times V \). In an {undirected graph}, edges are treated as unordered pairs.
A graph is bipartite if its nodes can be partitioned into two disjoint sets such that no edge connects nodes within the same set.
We restrict attention to undirected bipartite graphs. A graph $G$ can equivalently be represented by a variety of matrices, including the {adjacency matrix} \( A \in \mathbb{R}^{n \times n} \),
the graph {Laplacian} \( \ccalL \in \mathbb{R}^{n \times n} \) 
and their normalizations. We say the Laplacian is size-normalized when divided by the size of its edge set.


Given a graph \( G = (V, E) \), its line-graph \( L(G) = (V', E') \) is defined such that \( V' \) consists of one node for each edge in \( E \), and \( E' \subseteq V' \times V' \) contains an edge between two nodes in \( V' \) if and only if their corresponding edges in \( E \) share a common endpoint in \( G \).  

\label{prelims}

\subsection*{Square Degree Graphs}
The square-degree property characterizes sparse graph sequences with strong degree heterogeneity. As we discuss below, this condition is both necessary and sufficient for the associated line-graph sequence to be dense, which is the key structural mechanism underlying our convergence results.

A sequence of undirected graphs $\{G_n\}$, $G_n = (V_n, E_n)$, exhibits the {square-degree property} if there exists $c>0$ and $N_0 \in \mathbb{N}$ such that for all $n \geq N_0$,
\begin{equation}
    \sum_i \deg(v_{i,n})^2 \geq c \bigg( \sum_i \deg(v_{i,n}) \bigg)^2
    \label{eq: sqdeg}
\end{equation}
where $v_{i,n} \in V_n$ \citep{kandanaarachchi2024graphons}. 

A sequence of graphs $\{G_n\}$ that satisfies the square-degree property is {sparse} in that the growth of the number of edges $m$ with the number of nodes $n$ satisfies
\begin{equation}
    \lim_{n \to 
    \infty} \frac{m}{n^2} = 0 \text{.}
    \label{sparse}
\end{equation}

\begin{figure}[t!]
\centering
    \begin{subfigure}[t]{0.2\textwidth}
        \centering
        \includegraphics[height=.7in]{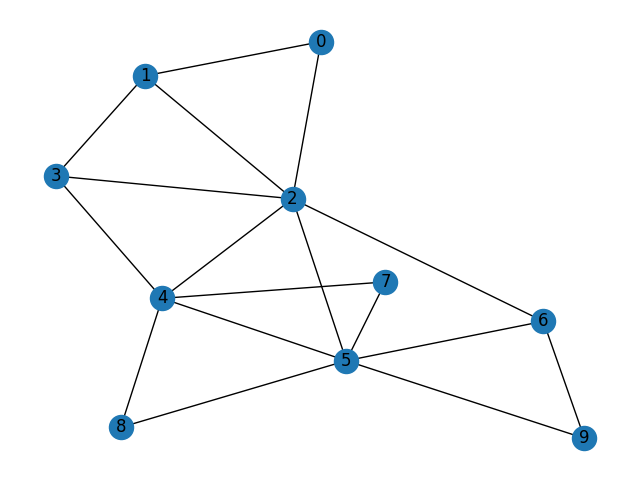}
        \caption{Superlinear BA \\($\alpha = 1.5, m=2)$}
        \label{sample_BA}
    \end{subfigure}%
    ~ 
    \begin{subfigure}[t]{0.2\textwidth}
        \centering
        \includegraphics[height=.7in]{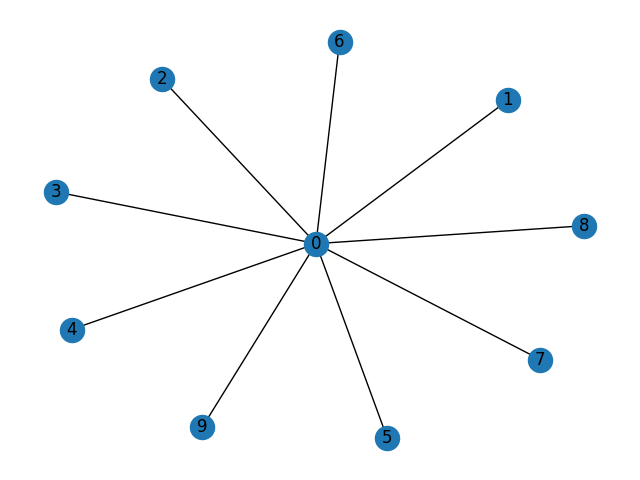}
        \caption{Star \\$K_{\{1,10\}}$}
        \label{sample_star}
    \end{subfigure}
        ~ 
    \begin{subfigure}[t]{0.2\textwidth}
        \centering
        \includegraphics[height=.7in]{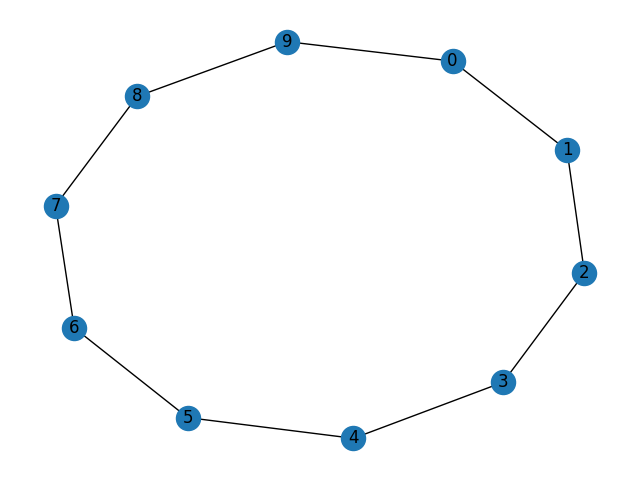}
        \caption{$10$-cycle}
        \label{sample_cycle}
    \end{subfigure}
        ~ 
    \begin{subfigure}[t]{0.2\textwidth}
        \centering
        \includegraphics[height=.7in]{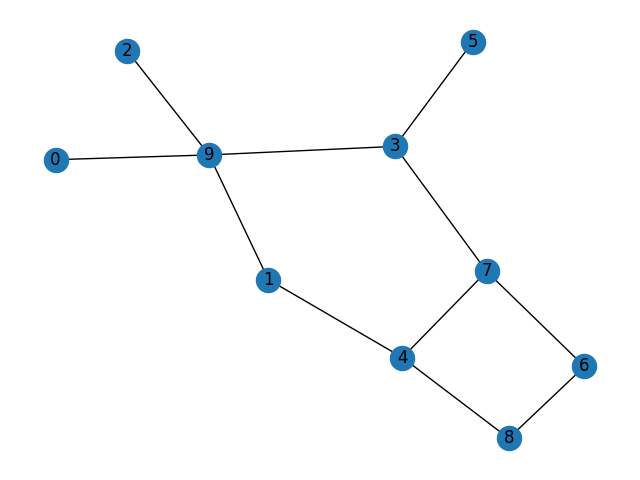}
        \caption{Erd\H{o}s-R\'{e}nyi\\ ($p = \frac{3}{n}$)}
        \label{sample_ER}
    \end{subfigure}
\caption{Examples of sparse graphs.}\label{graph_samples}
\end{figure}

This limit is strictly greater than zero for a sequence of dense graphs. Figures \ref{graph_samples}a--b show sparse graphs that satisfy the square-degree property, while Figures \ref{graph_samples}c--d show sparse graphs that do not.

We define the normalized degree profile for graphs satisfying the square degree property \eqref{eq: sqdeg}, which captures the asymptotic contribution of high-degree vertices to the total edge mass. In particular, if $\{G_n\}$ has degrees $d_{1,n} \geq d_{2,n} \geq \dots$, the normalized degree profile is 
\begin{equation}
\label{degprofile}
p^{(n)} = \bigg(\frac{d_{1,n}}{|E_n|}, \frac{d_{2,n}}{|E_n|}, \dots \bigg).
\end{equation}
We say the degree profile converges if $p^{(n)} \to p = (p_1, p_2, \dots)$ in $\ell^2$.

A property of square-degree graphs that will be key in the derivation of our spectral convergence results is that their line-graphs form a dense graph sequence. The sequence of line-graphs associated with a graph sequence $\{G_n\}$ will be denoted $\{L(G_n)\}$, with $L(G_n) = (V_n', E_n')$. In contrast with  \eqref{sparse}, $|E_n'|/|V_n'|^2$ has limit strictly greater than zero, defining the line graph as dense. The following lemma, adapted from \citep[Lemma 3.6]{kandanaarachchi2024graphons}, shows that the square-degree property is indeed both a necessary and sufficient condition for the line-graph sequence to be dense.
\begin{lemma}
Let $\{G_n\}$ be a sparse graph sequence and $\{L(G_n)\}$ be the corresponding line-graph sequence. Then, $\{G_n\}$ satisfies the square-degree property $ \iff \{L(G_n)\}$ is a dense graph sequence.
\label{sqdensity}
\end{lemma}


\label{sq_graphs}

\subsection*{Graphons}
{Graphons} are bounded symmetric measurable functions $\mathbf{W}: [0,1] \times [0,1] \to [0,1]$ that can be viewed as continuous analogues of large dense graphs.
As $\bbW$ is bounded and symmetric, it defines a self-adjoint integral linear operator $T_{\bbW}$ on $L^2([0,1])$. Specifically, given a function $X \in L^2([0,1])$ and variables \(u,v\in[0,1]\) representing node indices,
\begin{equation} \label{eqn:graphon_op}
(T_{\bbW} X)(v) = \int_{0}^1 \bbW(u,v) X(u) \text{d} u.
\end{equation} 

Graphons admit two standard interpretations: generative random graph models and limit objects of convergent dense graph sequences \citep{lovasz2012large,borgs2008convergent}. In the generative view, $\mathbf{W}$ defines a random graph by assigning each node a latent variable $u_i \in [0,1]$ and connecting node pairs with probability $\mathbf{W}(u_i,u_j)$. In the limit view, graphons arise as limits of dense graph sequences, where appropriately normalized graph statistics converge to their graphon counterparts. Under suitable node relabeling, this is equivalent to convergence in the cut norm. For a kernel $\bbK : [0,1]^2 \to [-1,1]$, the cut norm is defined as
\begin{equation}
\|\bbK\|_\square
\;\coloneqq\;
\sup_{S,T \subseteq [0,1]}
\left|
\int_{S \times T} \bbK(u,v)\,\mathrm{d}u\,\mathrm{d}v
\right|.
\label{eq:cutnorm-graphon}
\end{equation}
These convergence viewpoints enable asymptotic analysis of large dense graphs \citep{lovasz2012large,borgs2008convergent}.

We will also utilize graphons induced by finite graphs. Every undirected graph $G$ admits an induced graphon $\mathbf{W}_G$. The {induced graphon} is created by constructing a regular partition 
$I_j = [(j-1)/n, j/n), 1 \leq j \leq n-1$ and $I_n = [(n-1)/n, 1]$ and defining 
\begin{equation}
\mathbf{W}_G(u,v)=\sum_{j=1}^n\sum_{k=1}^{n}[A]_{jk}\mathbb{I}(u \in I_j)\mathbb{I}(v \in I_k).
\label{inducedgraphon}
\end{equation}

\label{graphons}

\subsection*{Line-Graphon Limits of Square Degree Graphs}
The square-degree property \eqref{eq: sqdeg} identifies sparse graph sequences whose edge interactions are sufficiently concentrated to induce dense structure after a line-graph transformation. This allows such sparse graphs to be embedded into the dense graph limit framework.

\begin{figure*}[!t]
\centering
\includegraphics[width=0.32\linewidth]{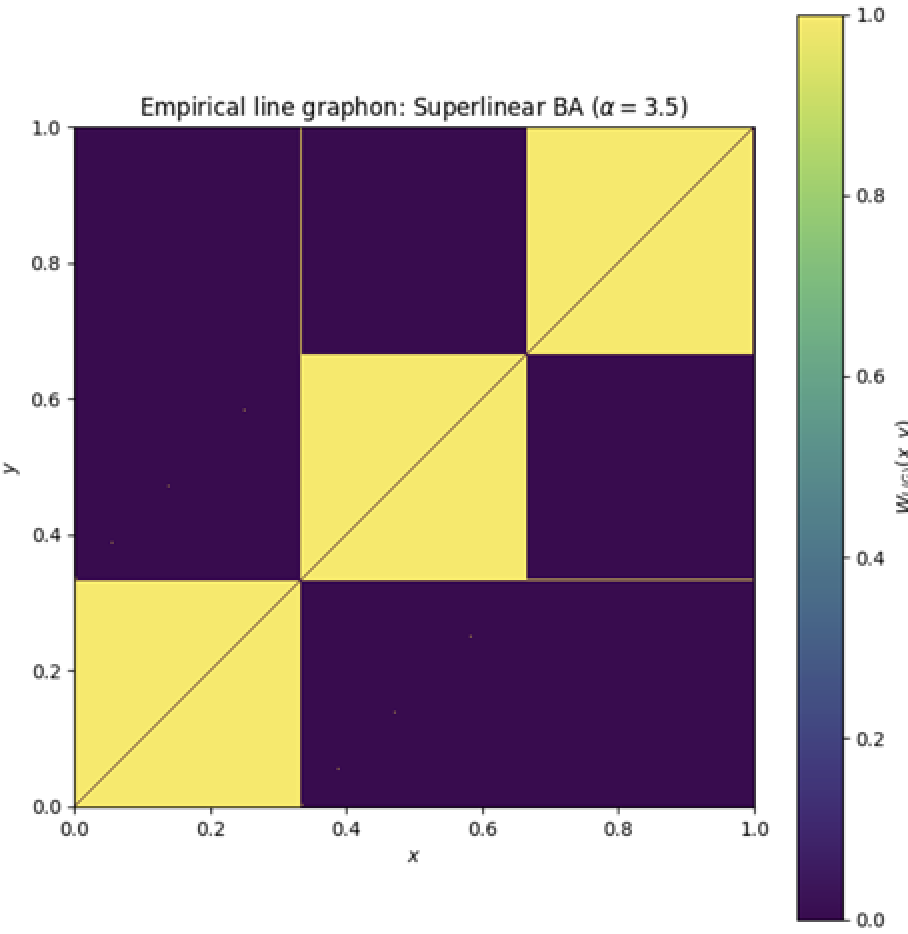}\hfill
\includegraphics[width=0.32\linewidth]{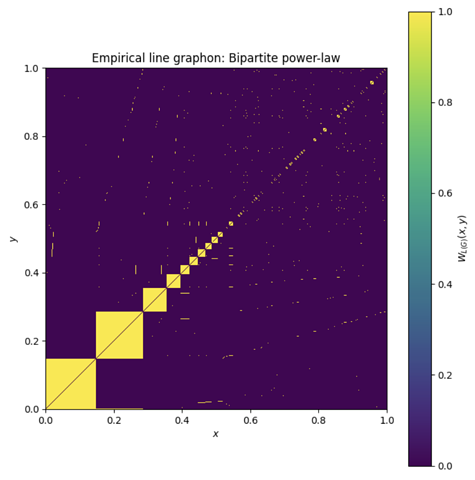} \hfill
\includegraphics[width=0.32\linewidth]{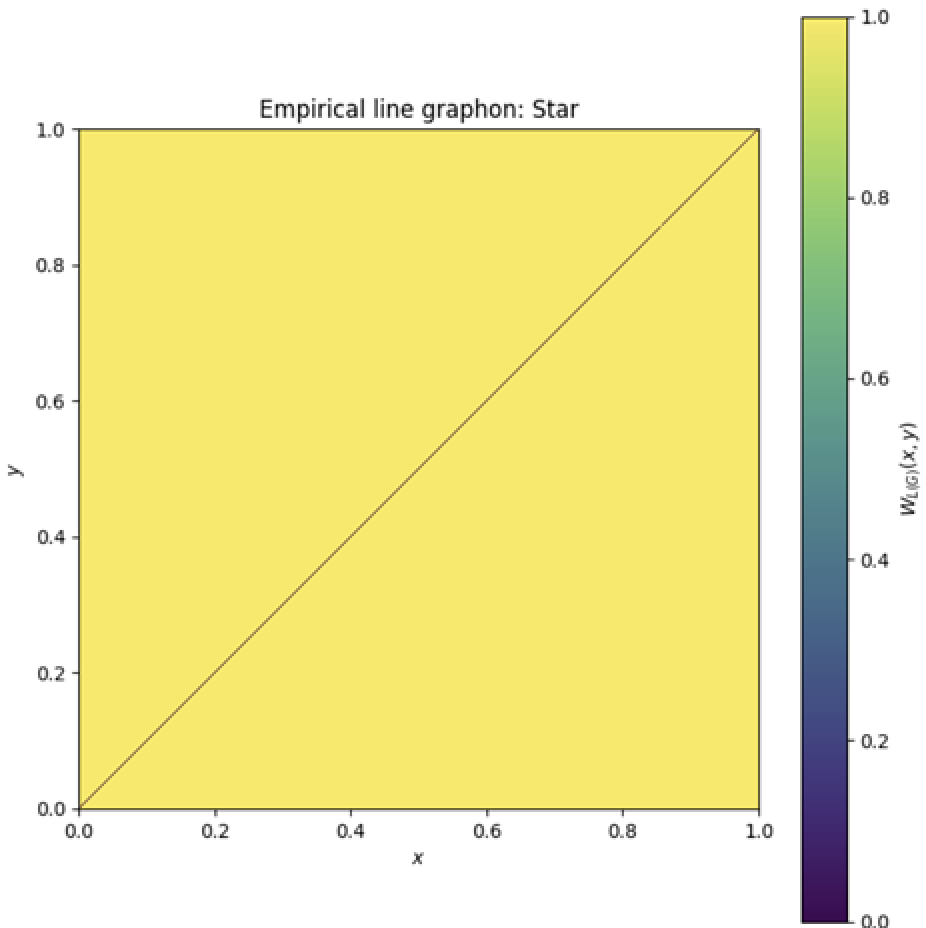}
\caption{
Empirical line-graphons derived from $1,000$ node graphs.
}
\label{fig:line_graphon}
\end{figure*}

Lemma \ref{sqcutnorm}, adapted from \citep[Lemma 3.10]{kandanaarachchi2024graphons}, shows that line-graph sequences associated with square-degree graph families admit nontrivial graphon limits, enabling meaningful spectral analysis via the operator \eqref{eqn:graphon_op}. This nontrivial limiting graphon is the {line-graphon}. Visualizations can be found in Fig. \ref{fig:line_graphon}.

\begin{lemma}
Let $\{G_n\}$ satisfy the square-degree property and $\{L(G_n)\}$ be the corresponding line-graph sequence. If $\{L(G_n)\}$ converges to a graphon $\mathbf{U}$, then $\mathbf{U}$ has a strictly positive cut-norm.
\label{sqcutnorm}
\end{lemma}

\label{line_graphons}

\section*{Spectral Convergence of Sparse Bipartite Graphs}
\label{main}
A central obstacle in spectral analysis of sparse graphs is the lack of pure point eigenvalue convergence under graph growth. Even in weakly-convergent sparse graph sequences, Laplacian spectra typically do not stabilize at the level of individual eigenvalues \citep{Salez2020SpectraSparse}. 

In this section, we show that this obstruction can be overcome for a class of sparse bipartite graphs by exploiting two important structural properties: (i) a spectral correspondence between bipartite graph Laplacians and line-graph adjacency operators, and (ii) the square-degree property \eqref{eq: sqdeg}, which ensures the associated line-graph sequence is dense. Together, these properties reduce spectral analysis of sparse bipartite graphs to classical operator convergence results for dense graph sequences.


\subsection*{Spectral Correspondence between Bipartite Laplacian and Line Graph Adjacency}
The reduction is enabled by a direct spectral correspondence between bipartite graphs and their associated line-graphs. This is formally stated in Lemma \ref{eig}.


\begin{lemma} \label{eig}
Let $G = (V, E)$ be a bipartite graph with Laplacian eigenvalues $\lambda_0 = 0 \leq \lambda_1(\ccalL) \leq \lambda_2(\ccalL) \leq \dots \leq \lambda_{n-1}(\ccalL)$. Then, the eigenvalues of the adjacency matrix of its line-graph satisfy $\lambda_0(A_{L(G)}) = -2$ with multiplicity $|E|-n + 1$, and $\lambda_i(A_{L(G)}) = \lambda_i(\ccalL) - 2$ for $1 \leq i \leq n - 1$. 
\end{lemma}

This spectral correspondence allows eigenvalue-level statements for sparse bipartite graphs to be transferred to their dense line-graph representations. The result follows from standard relationships between the Laplacian and the line-graph adjacency via the incidence matrix \eqref{B}. A detailed proof and numerical analysis are provided in Apps. \ref{isospectral} and \ref{eigalign}, respectively.


Throughout the remainder of the paper we index eigenvalues in nonincreasing order ($\lambda_1 \geq \lambda_2 \geq \dots $). Laplacian eigenvalues in Lemma \ref{eig} retain the standard ascending ordering ($0 = \lambda_0 \leq \lambda_1 \leq \lambda_2 \leq \dots$). 

\label{eval_relationship}

\subsection*{Spectral Convergence via Line-Graph}

The square-degree property \eqref{eq: sqdeg}, together with the degree-profile condition \eqref{degprofile} and Lemma~\ref{eig}, yields eigenvalue convergence for size-normalized Laplacians of sparse bipartite graphs. Lemma \ref{degtographon} formalizes the restriction to such graph sequences. At a high level, vertices induce cliques in the line-graph whose sizes are determined by their degrees; a small number of high-degree vertices dominate, and the limiting structure is determined by the degree profile, yielding a disjoint-clique graphon. This follows from line-graph limit results in \citep{kandanaarachchi2024graphons} together with the characterization of clique-based graphons via mass partitions \citep{janson2016graph}. Full details are provided in App.~\ref{lem4}.

\begin{lemma}
\label{degtographon}
Let $\{G_n\}$ be a sequence of bipartite graphs satisfying the square-degree property \eqref{eq: sqdeg}. Suppose the normalized degree profile \eqref{degprofile} converges to $p = (p_1, p_2, \dots)$ in $\ell^2$. Then the associated line-graph sequence $\{L(G_n)\}$ converges in the cut distance \eqref{cutdistance} to a graphon $U_p$, which is (up to measure-preserving equivalence) a disjoint-clique graphon determined by the limiting clique profile $p$.
\end{lemma}

Leveraging the above result along with norm equivalences for graphon operators, we are able to show spectral stability for square-degree bipartite graphs, which is the main result of this paper. 

\begin{theorem}\label{laplacesparse}
Let $\{G_n\}$ be a sequence of bipartite graphs satisfying the square-degree property \eqref{eq: sqdeg} with convergent normalized degree profile \eqref{degprofile}. Let $G_{n_1}=(V_{n_1}, E_{n_1})$ and $G_{n_2}=(V_{n_2}, E_{n_2})$ be graphs from this sequence with size-normalized Laplacians $\ccalL_{n_1}$ and $\ccalL_{n_2}$, respectively. Let $\lambda_i(\ccalL_n)$ denote the $i^{th}$ largest eigenvalue of $\ccalL_n$. Then, as $n_1, n_2 \to \infty$, for each $i\geq1$,
\begin{equation}
    |\lambda_i(\ccalL_{n_1}) - \lambda_i(\ccalL_{n_2})| \to 0 .
\end{equation}

\end{theorem}


Theorem~\ref{laplacesparse} establishes eigenvalue-level convergence 
for fixed sequences of square-degree graphs. The spectral ordering considered here corresponds, via Lemma~3, to the dominant eigenvalues of the normalized line-graph adjacency operators, and hence aligns with the pure-point spectrum of the limiting graphon operator. The proof follows directly 
from convergence of the line-graph sequence to its limit graphon, combined with the invariance 
of graphon operator eigenvalues under measure-preserving bijections.

In particular, Theorem \ref{laplacesparse} identifies a structural regime in which sparsity 
remains compatible with pure point spectral convergence, a setting where 
existing results generally apply only at the level of spectral measures 
or empirical distributions \citep{Salez2020SpectraSparse}.


\label{spectral_bound}

\subsection*{Proof Sketch of Theorem \ref{laplacesparse}}
The proof proceeds by mapping the Laplacian spectrum of a sparse bipartite graph to the adjacency spectrum of its line-graph via Lemma~\ref{eig}. The square-degree property \eqref{eq: sqdeg} and normalized degree profile \eqref{degprofile} ensure the resulting line-graph sequence is dense and convergent. At the line-graph level, convergence is quantified by comparing the spectra of the induced line-graphons ~\eqref{inducedgraphon} with that of the limiting line-graphon. This comparison is controlled via a Weyl-type inequality for compact self-adjoint operators, formalized in Prop.~\ref{weyl's}.

\begin{proposition}[Weyl-type inequality for graphon operators] \label{weyl's} 
    Let \( \mathbf{W} \) be a graphon and let \( G_n \) converge to \bbW with induced graphon \({\mathbf{W}_{n}} \) ({\ref{inducedgraphon}}) and associated graphon operator \(T_{\mathbf{W}_{n}} \) \eqref{eqn:graphon_op}. Then, for all $1 \leq i \leq n$,
    \begin{align} \label{eq: weyl's} 
    |\lambda_i(\mathbf{W})-\lambda_i({\mathbf{W}_{n}})| \leq \|T_\mathbf{W} - T_{\mathbf{W}_{n}}\|_2.
    \end{align}
\end{proposition}

To bound the operator norm on the right-hand side of \eqref{eq: weyl's}, we use a standard norm equivalence relating the operator norm of an integral operator to the cut norm \eqref{eq:cutnorm-graphon} of its kernel. This result is adapted from \citep[Prop.~4]{ruiz2020graphonsp} and summarized in Prop.~\ref{bound}.

\begin{proposition} \label{bound}
Let $G_n$ be a graph converging to a graphon $\mathbf{W}$ and let $\mathbf{K} = {\mathbf{W}_{n}} - \mathbf{W}$ where ${\bbW_n}$ is the graphon induced by $G_n$. Noting that $\mathbf{K}: [0,1] \times [0,1] \rightarrow [-1,1]$, we have
    \begin{align}
    ||\mathbf{K}||_{\Box} \leq ||T_{\mathbf{K}}||_{2} \leq \sqrt{8||\mathbf{K}||_{\Box}} \text{.}
    \end{align}
\end{proposition}

Next, lemma \ref{deterministic} below establishes that graphon operator eigenvalues are invariant under measure-preserving bijections, enabling cut distance \eqref{cutdistance} convergence to be transferred to eigenvalue (operator-level) convergence. The proof is deferred to App. \ref{determ_proof}.

\begin{lemma}\label{deterministic}
Let $\mathbf{W}$ be a graphon and $\phi \in \Phi$ a measure-preserving 
bijection on $[0,1]$. Then $T_{\mathbf{W}}$ and $T_{\mathbf{W}^\phi}$ 
are unitarily equivalent and thus isospectral:
\[
\lambda_i(T_{\mathbf{W}^\phi}) = \lambda_i(T_{\mathbf{W}}) 
\quad \text{for all } i.
\]
\end{lemma}

Combining Propositions~\ref{weyl's} and \ref{bound} with Lemma~\ref{deterministic} yields eigenvalue convergence for induced line-graphons, stated in Lemma~\ref{lgbounds}. The proof is provided in App. \ref{lem3.6}.

\begin{lemma} \label{lgbounds}
Let $\{G_n\}$, $G_n = (V_n, E_n)$, be a sequence of graphs that satisfies the square-degree property \eqref{eq: sqdeg}, and let $\{L(G_n)\}$ be the associated line-graph sequence converging to a graphon $\mathbf{U}$. Let ${\mathbf{U}_{n}}$ be the graphon induced by $L(G_n)$. Then, as $n \to \infty$, for all $i$
\begin{equation}
|\lambda_i({\mathbf{U}_{n}}) - \lambda_i({\mathbf{U}})| \to 0.
\end{equation}

\end{lemma}

The argument reduces spectral comparison for a regime of sparse bipartite graphs to their dense line-graph representations. By Lemma~\ref{eig}, Laplacian eigenvalue differences correspond to differences in the line-graph spectra. Lemma~\ref{lgbounds} then establishes convergence of the induced line-graphon spectra to that of the limiting line-graphon \(\bbU\). Applying this convergence result to both graphs and using the triangle inequality yields the stated eigenvalue asymptotics for the original sparse Laplacians. A complete proof is provided in App.~\ref{thm1_proof}.

\label{bound_proof}


\section*{Case Study: Superlinear Preferential Attachment Graphs}
\label{case study}
SPA trees, a well-studied class of sparse bipartite graphs with extreme degree heterogeneity and an explicit infinite limit object, provide an illustrative example of the pure point spectral convergence result of Theorem \ref{laplacesparse}. 

\subsection*{SPA Trees}
Preferential attachment (PA) graphs are generated sequentially by attaching new nodes to existing nodes with probability proportional to a function of their degree.
Under the Barab\'{a}si-Albert (BA) model, the process initializes with a small connected graph consisting of $m_0 > m$ nodes, with $m \geq 2$. At each time step, a new node is added along with $m$ edges, each of which connects to an existing node $i$ with probability
\begin{equation} \label{eq:BA}
 \Pi(d_i) = \frac{d_i^\alpha}{\sum_j d_j^\alpha}   
\end{equation}
where $d_i$ is the degree of node $i$ and $\alpha$ is the attachment exponent \citep{albert2002statistical}.
When $\alpha>1$ the model yields an SPA graph exhibiting hub condensation. For SPA trees, the model initializes with a two-node tree with one edge, and $m=1$. 
SPA trees, which have been studied extensively in the probability literature \citep{oliveira2005connectivity,sethuraman2016growth,szabo2002shortest,pittel1994note}, satisfy the square-degree property \eqref{eq: sqdeg}; moreover, hub condensation in the superlinear regime induces a well-defined ordered edge-normalized degree profile.


\label{spa_trees}

\subsection*{SPA Limits and Their Line Graphons}
The SPA tree limit object is defined as follows. Let $T_n$ denote rooted trees with at most $n$ nodes, and let $a = \lfloor \frac{\alpha}{\alpha - 1} \rfloor$. The limit consists of a finite core tree with root $v$, to which infinite copies of $T$ are attached for each $T \in T_a$. This yields infinitely many nodes of degree $k$ for $1 \leq k \leq a$, while only the root $v$ may have degree exceeding $a$ \citep{oliveira2005connectivity}. We refer to this object as $SPA_{\mathrm{lim}}$. As we show in the following, this construction enables an operator-theoretic characterization of SPA beyond finite-graph spectral comparisons (as in Theorem~\ref{laplacesparse}), allowing to identify the limiting operator and to relate its spectrum to that of the associated line-graphon.

The key first step is characterizing the behavior of the associated line-graphs. In $SPA_{\mathrm{lim}}$, a single high-degree hub dominates the structure while lower-degree components remain finite. Under the line-graph mapping, stars centered at high-degree vertices induce large cliques, so the resulting line-graphs converge to a nontrivial graphon reflecting this clique-dominated structure.

\begin{proposition} \label{SPAlinelimit}
Let $SPA_m$ be a sequence of finite SPA trees converging locally to $SPA_{\mathrm{lim}}$. Then the associated line-graph sequence $L(SPA_m)$ converges to a graphon $\bbW_{SPA_{\mathrm{lim}}}$.
\end{proposition}

The proof follows from the explicit structure of $SPA_{\mathrm{lim}}$ and the fact that stars induce cliques under the line-graph construction. Details are provided in App. \ref{linegraph}.
\label{spa_limits}

\subsection*{Limit Laplacian/Line-Graphon Correspondence}
We construct a regularized self-adjoint Laplacian operator $\ccalL_\infty$ on $SPA_{\mathrm{lim}}$
via the adjacency matrix of an $m$-edge SPA tree, denoted $SPA_m$, as follows. 

Enumerate the edges incident to $r$ as $e_k=(r,y_k), k \geq 1$. Assign edge $e_k$ weight $w^m_k$ such that $\sum_k w^m_k < + \infty$ for finite $m$ and $\lim_{m\to\infty} w_k^m = w_k$ with $(w_k) \in \ell^1$. Define
\[A^m_w(x,y) = \begin{cases}
    w^m_k & \text{if } x=r, \ y=y_k \\
    1 & \text{if } x \sim y, \  x,y \neq r \\
    0 & \text{otherwise.} 
\end{cases}\]
Then, the Laplacian associated with $A^m_w$ is
\begin{align}
\ccalL^m_wf(y) = \sum_{y \sim x} A^m_w(y,x)[f(y) - f(x)]
\end{align}
for an arbitrary function $f$ on the nodes of $SPA_m$.

Prop.~\ref{limitlap} shows that such weights exist and yield a well-defined, bounded,
self-adjoint limit Laplacian operator. The proof is deferred to App.~\ref{laplimit}.

\begin{proposition}
Let $V(T_\infty)$ be the vertices of $SPA_{lim}$. Then, $\ccalL_\infty = \lim_{m \to \infty}\frac{\ccalL^m_w}{m}$ is a well-defined, bounded, self-adjoint operator on $\ell^2(V(T_\infty))$.
\label{limitlap}
\end{proposition}

This construction provides an operator-theoretic description of the SPA limit independent of the line-graphon framework, allowing for comparison between classical sparse graph limit operators and the graphon operators obtained via dense line-graph representations. It can be shown that this Laplace operator has trivial bulk spectra, as stated in Lemma \ref{bulkspectra} and proved in App. \ref{bulkproof}.

\begin{lemma}\label{bulkspectra}
    The bulk empirical spectral measure of $\ccalL_\infty$ converges weakly to $\delta_0$.
\end{lemma}

Having established the existence of a limiting Laplacian, we investigate if the spectral correspondence between bipartite Laplacians and line-graph adjacency operators established by Lemma \ref{eig} persists in the infinite setting for SPA trees. The following lemma provides a confirmation. The proof is deferred to App. \ref{limeigproof}.

\begin{lemma} \label{limiteig}
Consider the limit size-normalized Laplacian $\ccalL_\infty = \lim_{m\to\infty}{\ccalL^m_w/m}$ and let $\{\lambda_i(\ccalL_\infty)\}_{i}$ be its eigenvalues. 
Then, for each fixed index $i$,
\[
\lambda_i\!\big(T_{\mathbf{W}_{SPA_{\mathrm{lim}}}}\big)
=
\lambda_i(\ccalL_\infty).
\]
\end{lemma}

This establishes the extension of the spectral correspondence to the infinite SPA limit. Note that the nonzero eigenvalue shift by 2 seen in the finite case (Lemma \ref{eig}) vanishes as $m \to \infty$.

The final step is verifying that $\mathcal{L}_\infty$ indeed has a pure point spectrum. 

\begin{proposition}
    $\ccalL_\infty$ has pure point spectrum. 
\end{proposition}
\begin{proof}
    By Lemma \ref{bulkspectra} the only non-zero eigenvalues are outliers. By Lemma \ref{limiteig} these outliers are eigenvalues of a compact operator and thus pure point.
\end{proof}

\label{spa_limit_laplacian}

\subsection*{Spectral Convergence to the Infinite Limit}
Utlizing the above ingredients—existence of a limiting Laplacian, isospectrality with the associated line-graphon, and convergence guarantees for dense line-graph sequences—we obtain eigenvalue-level convergence of finite SPA Laplacians to an explicit infinite-dimensional operator.

\begin{theorem} \label{SPAcor}
Let $\{G_n\}$ be a sequence of SPA trees that converge to $SPA_{lim}$ with limit size-normalized Laplacian $\ccalL_\infty = \lim_{m\to\infty}{\ccalL^m_w/m}$. Let $G_{n}=(V_{n}, E_{n})$ be a sample from this sequence with size-normalized Laplacian $\ccalL_{n}$. Then as $n \to \infty$, for all $i$,
\begin{equation}
    |\lambda_i(\ccalL_{n}) - \lambda_i(\ccalL_\infty)| \to 0.
\end{equation}
\end{theorem}
\begin{proof} 
$\{G_n\}$ has corresponding line-graph sequence,  $\{L(G_n)\}$ converging to ${\mathbf{W}_{SPA_{lim}}}$ (Prop. \ref{SPAlinelimit}). Let the induced graphon of $L(G_{n})$ be ${\mathbf{W}_{n}}$. By Lemmas \ref{eig},  
\ref{limiteig}, and \ref{lgbounds}, as $n \to \infty$, for all $i$,
    \begin{align*}
    |&\lambda_i(\ccalL_{{n}}) - \lambda_i(\ccalL_\infty)| \\
    &=|\lambda_i({\mathbf{W}_{n}}) + \frac{2}{|E_n|} - \lambda_i({\mathbf{W}_{SPA_{lim}}})|
    \to 0.
    \end{align*}
\end{proof}
Note that, unlike in the general results of Sec. \ref{line_graphons}, the limit in this case is explicitly identified. This case study demonstrates that the line-graphon framework is not merely an abstract tool for comparing finite sparse graphs, but is also consistent with infinite-limit constructions when such limits exist. In SPA trees, the dense line-graph representation bridges finite graphs and their infinite counterpart, yielding spectral convergence to a well-defined limiting Laplacian operator.

This example shows that the framework developed in this paper unifies two regimes that are often treated separately---abstract spectral convergence results for sparse graphs and explicit operator limits arising in denser, often random graph models. The square-degree property plays a central role in this unification, enabling dense representations that faithfully capture spectral behavior even in highly sparse settings.

\label{spa_infinite_limit}

\section*{Numerical Results}
We present two numerical experiments that empirically support Theorems~\ref{laplacesparse} and~\ref{SPAcor}, and an illustrative experiment investigating the stability of learned GNN representations across graph sizes within sparse bipartite graph families. Full experimental settings and other supporting experiments are stated in App. \ref{compute} and App. \ref{addexp}, respectively.

\label{num_res}

\subsection*{Eigenvalue Convergence on Synthetic Networks}
This experiment samples graphs from sparse bipartite graph sequences satisfying the square-degree property \eqref{eq: sqdeg} and exhibiting heavy-tailed degree structure (consistent with stable normalized degree profiles under graph growth) to study Laplacian eigenvalue convergence. The graph families considered are: Star and SPA with $\alpha=3.5$ and $\alpha=5$. Additionally, it uses samples from a $3$-regular bipartite graph as a comparison point. 

To quantify convergence, we compute the mean absolute error over the three largest eigenvalues of the size-normalized Laplacian. For each graph family, we compare the eigenvalues of each sampled graph with those of the largest graph in the sequence. Errors are averaged across independent runs, and shaded regions indicate one standard deviation. The resulting normalized errors are plotted against graph size in Fig.~\ref{fig:num_res} (left). Across graph families satisfying the model assumptions, eigenvalue differences decrease as graph sizes increase, indicating stabilization consistent with Thm.~\ref{laplacesparse}. Furthermore, while the samples from the regular graph converge, it is to a much larger error, providing support for the importance of the assumptions used in Thm.~\ref{laplacesparse}.

\label{synthetic_ex}

\subsection*{Eigenvalue Convergence in Real-World Graphs}
This experiment samples subgraphs from real-world sparse bipartite networks to study Laplacian eigenvalue convergence under graph growth. We evaluate four bipartite interaction datasets: MovieLens-1M \citep{harper2015movielens}, Goodreads Goodbooks-10k \citep{goodbooks10k}, Amazon Fine Food Reviews \citep{mcauley2013hidden}, and Book-Crossing \citep{ziegler2005improving}, represented as user--item graphs with edges corresponding to observed interactions. Although these datasets are not constructed to satisfy the square-degree property \eqref{eq: sqdeg} or exhibit stable normalized degree profiles \eqref{degprofile}, they demonstrate similar sparsity and heavy-tailed degree distributions, empirical validation of which can be found in App. \ref{realsq}. This experiment evaluates whether the predicted spectral convergence behavior extends beyond idealized models.

\begin{figure*}[!t]
\centering
\includegraphics[width=0.32\linewidth,height=3.7cm]{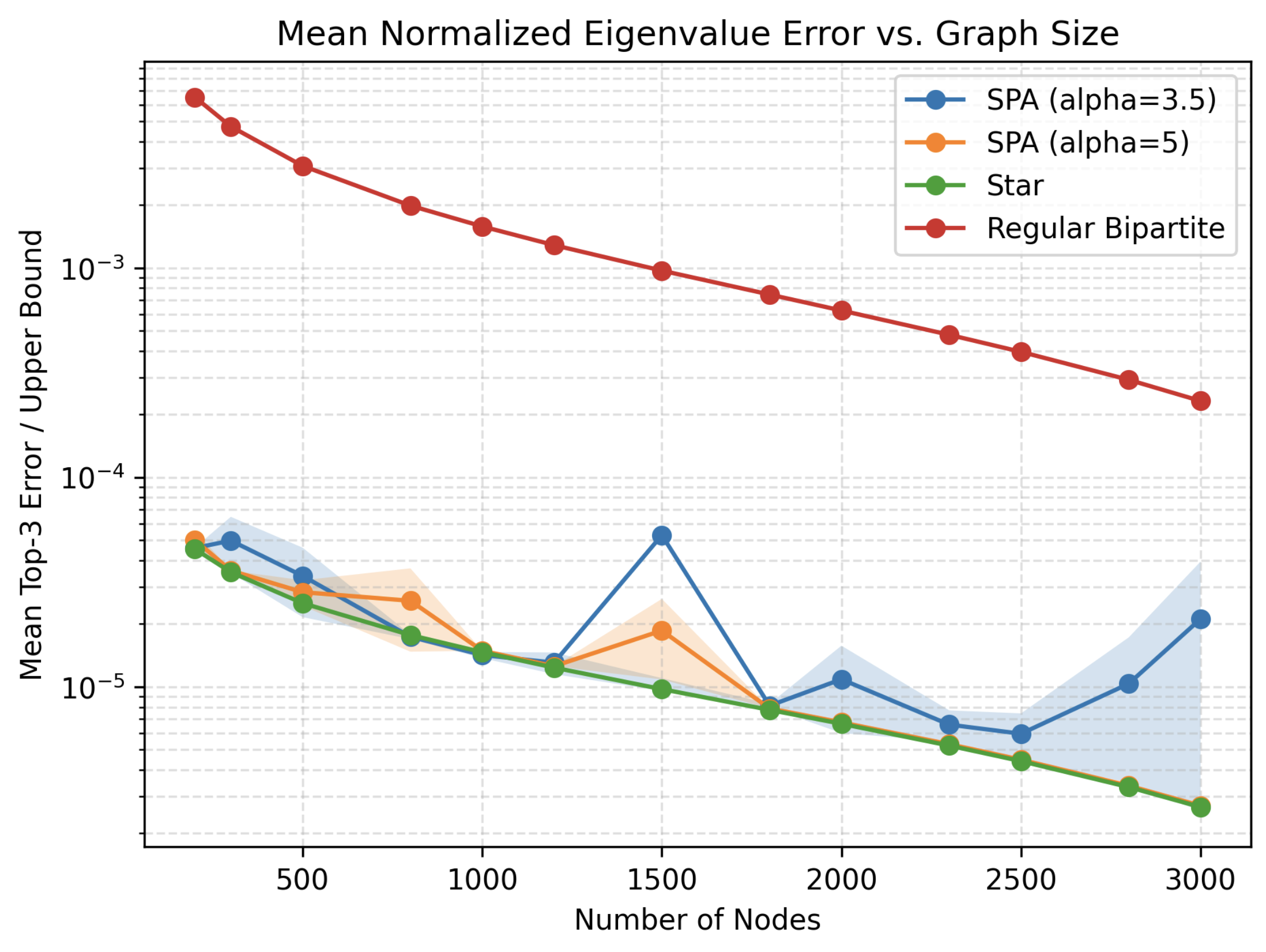}\hfill
\includegraphics[width=0.32\linewidth,height=3.7cm]{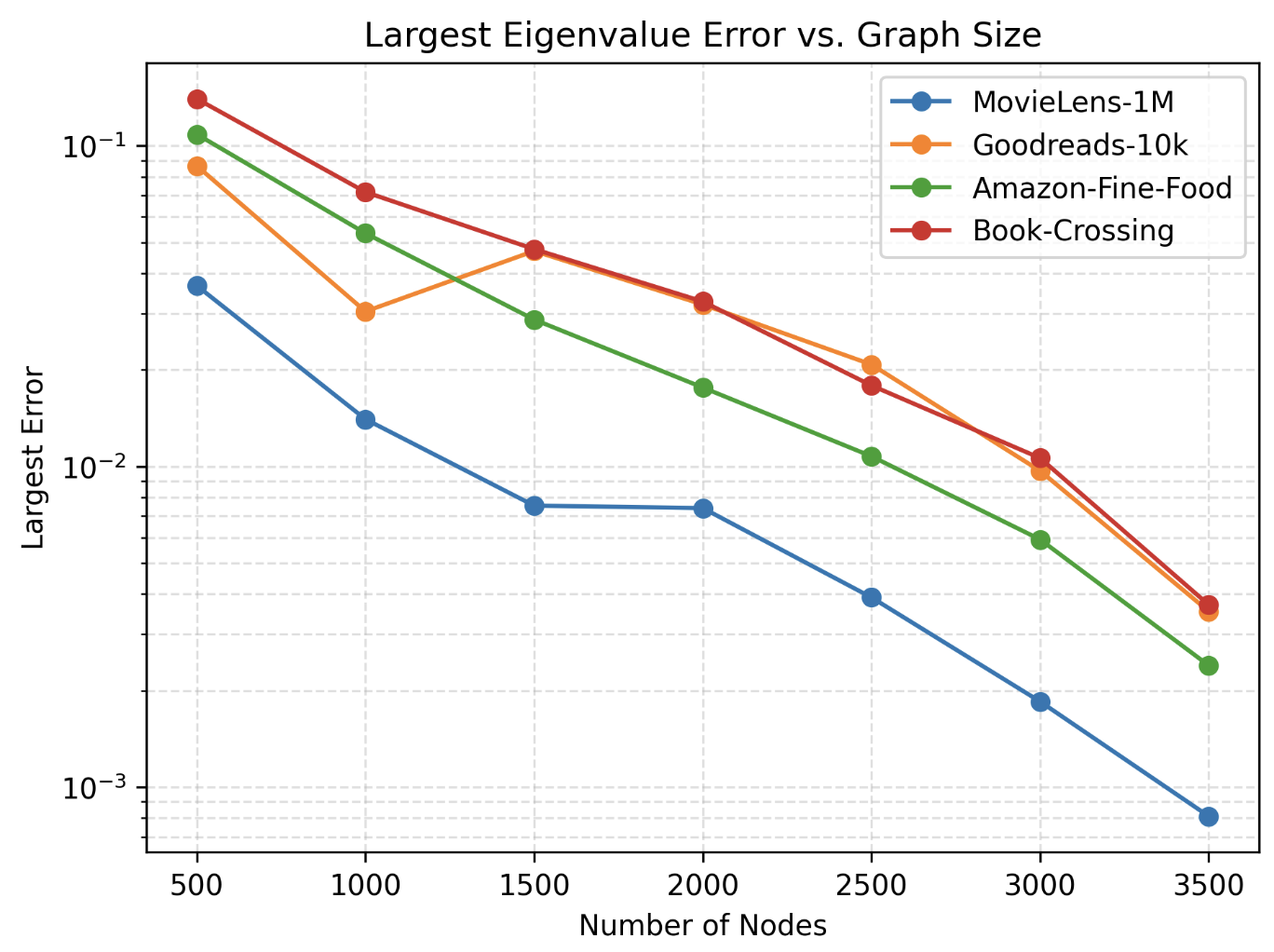} \hfill
\includegraphics[width=0.32\linewidth,height=3.7cm]{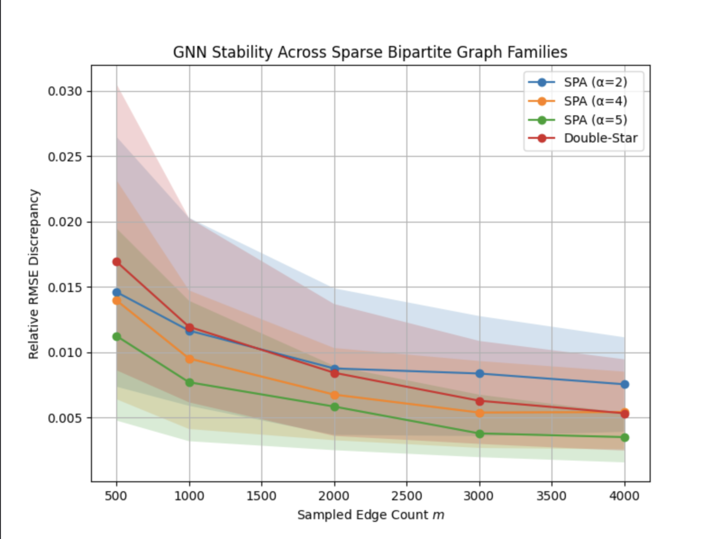}
\caption{
Numerical results. Left: Synthetic eigenvalue convergence. Center: Real-world eigenvalue convergence. Right: Relative RMSE discrepancy of a fixed GNN on edge-induced subgraphs; lower values indicate greater stability. Shaded regions indicate standard deviation for the first plot and IQR for the GNN plot.
}
\label{fig:num_res}
\end{figure*}

We compare the absolute differences between the largest eigenvalues of the size-normalized Laplacian of each sampled subgraph and those of the largest sampled subgraph. For each subgraph size, the maximum eigenvalue errors are plotted as a function of subgraph size, as shown in Fig.~\ref{fig:num_res} (center). Across all datasets, the eigenvalue differences decrease as subgraph size increases, indicating spectral stabilization consistent with Thm.~\ref{laplacesparse}.

\label{real_ex}

\subsection*{GNN Stability}
Convolutional GNNs can be interpreted as spectral operators \citep{gama2019stabilityscat, defferrard17-cnngraphs}, with layers corresponding to polynomial filters of graph operators. As transferability of GNNs across graphs depends on the stability of these operator's eigenvalues \citep{gama2020graphs, ruiz20-transf}, Theorem~\ref{laplacesparse} motivates investigating if graph families exhibiting stronger spectral stabilization also exhibit more stable learned representations under graph-size perturbations.

We consider the following graph families: SPA with $\alpha \in {2,4,5}$ and a balanced double-star bipartite graph. For each family, a graph neural network using the size-normalized Laplacian as its graph propagation operator is trained once on the full graph, using fixed random node features and supervision on $50\%$ of the nodes. The trained model is then evaluated without retraining on edge-induced subgraphs obtained by randomly sampling edges and retaining their incident nodes. For each sampled subgraph, the size-normalized Laplacian is recomputed using the sampled graph. Stability is measured using the relative discrepancy between the prediction error on a sampled subgraph and that on the full graph. This evaluates whether learned representations remain stable as the underlying size-normalized graph operator is perturbed through edge-induced restriction. Since stable behavior under graph perturbations is closely related to transferability across graph sizes, this experiment serves as an illustrative proxy for transfer stability within sparse graph families.

We emphasize that this experiment is illustrative rather than a direct empirical validation of Theorem~\ref{laplacesparse}. Although the GNN propagation operator is explicitly constructed from the same size-normalized Laplacian appearing in the theoretical analysis, the experiment concerns the stability of a particular learned nonlinear model and its prediction error, rather than direct convergence of Laplacian eigenvalues. Thus, the results provide empirical evidence for the relationship between stabilization of the graph operator under edge-induced sampling and transferability of learned representations, rather than a direct test of the theorem itself. For the corresponding experiment using a standard GCN with degree-normalized adjacency, see App. ~\ref{gcn_exp}.

\label{transferability}

\section*{Related Work}

A substantial body of work studies spectral convergence for dense and moderately sparse graph sequences through graph limit theory. Graphon convergence has been shown to imply convergence of associated spectral objects, including adjacency and Laplacian operators, under appropriate normalization \citep{borgs2008convergent, borgs2012convergent}. This framework links spectra of large graphs to compact integral operators induced by the limiting graphon. Related operator-theoretic formulations appear in the graph limit literature of Lovász and collaborators \citep{lovasz2006limits, lovasz2012large}. Related phenomena arise in geometric settings, where graph Laplacians constructed from random samples converge to Laplace--Beltrami operators on manifolds \citep{belkin2008towards, calder2022improved}. Across these settings, spectral convergence is driven by an underlying continuum operator limit.

In sparse graphs, operator-level convergence generally fails, and analysis instead focuses on empirical spectral distributions or local weak limits. A large body of literature studies limiting spectral measures of sparse random graphs and matrices using resolvent methods and local weak convergence. Under finite-moment degree assumptions, these methods establish convergence of spectral measures for sparse graph and configuration models; see \citet{bordenave2019lecture} for a comprehensive treatment. Unlike dense graphon limits, these results emphasize distributional convergence rather than eigenvalue-level convergence, reflecting the irregularity of sparse graph spectra. Our approach imposes a structural condition on the degree profile that controls global degree concentration, rather than local neighborhood structure. While local weak convergence (Benjamini–Schramm limits) describes the distribution of finite neighborhoods in sparse graphs, it typically yields convergence of spectral measures but not individual eigenvalues. Our results identify a complementary regime in which eigenvalue-level convergence can be recovered through a dense line-graph representation.

Several works study spectral properties of bipartite and nearly bipartite sparse graphs. \citet{kunegis2015exploiting} develops spectral methods tailored to bipartite networks, while \citet{cranston2020sparse} characterizes sparse graph families admitting nearly bipartite decompositions. From a random matrix perspective, \citet{Dumitriu_2014} derives limiting spectral distributions for bipartite biregular random graphs, and \citet{martinez2019spectral} analyzes eigenvector localization and spectral scaling in random bipartite models. More broadly, \citet{zhu2020graphon} uses graphon methods to study limiting spectral distributions of inhomogeneous sparse graphs. In contrast, our work establishes eigenvalue-level convergence guarantees for sparse bipartite graph sequences satisfying the square-degree property \eqref{eq: sqdeg} together with mild degree profile regularity by reducing the problem to dense line-graph operators. This yields a structural regime in which discrete spectral convergence can be recovered and connects these guarantees to applications such as GNN transferability.

\label{related_works}

\section*{Conclusion}\label{conclusion}
In this work, we studied spectral convergence of sparse bipartite graphs through the graphon limits of their line-graphs. We derived asymptotic convergence for Laplacian eigenvalues under structural assumptions on bipartiteness and well-behaved degree concentration. Our approach embeds certain sparse graph families into the dense graphon framework, recovering compactness and discrete spectral convergence, as supported by numerical experiments on synthetic and real-world data. Our results provide a principled framework for understanding GNN stability and transferability across graph sizes by reducing sparse spectral convergence questions to tractable dense graphon limits. More broadly, this work identifies a structural regime in which sparse graph spectra recover the discrete operator behavior typically associated only with dense graph limits.

A main limitation is that the convergence results do not include explicit rates in terms of graph size. Deriving such 
rates requires quantifying the rate of cut distance convergence of the 
deterministic line-graph sequence to its limit graphon, which is difficult without additional assumptions, unlike for a stochastic sequence where the graphon sampling lemma applies \cite[Lemma 10.15]{lovasz2012large}. Additionally, the 
structural assumptions, while satisfied by well-studied 
graph families such as star graphs and superlinear preferential attachment 
trees, restrict the theory to graphs with sufficiently 
concentrated and well-behaved degree structure.

Future work includes extending this analysis to sparse graphs that are perturbations of bipartite graphs \citep{cranston2020sparse}, incorporating graphon regularity assumptions to obtain quantitative convergence rates, 
and investigating end-to-end learning guarantees for GNNs under spectral stability. 

\subsection*{AI use statement}

In this work, we used generative AI tools for refining written proofs.
We have not used generative AI tools for synthetic dataset generation, development of theoretical models or conceptual frameworks, formulation of mathematical claims, determination of critical ingredients of mathematical claims, design or feedback on research methodology or experiments, implementation of methods, cleaning and reformatting datasets, support for qualitative and thematic data analysis, or interpretation of results. Assistance with translation is not applicable to this work.
Additionally, we used generative AI tools for modification of figures, editing software code, and improving readability.
We have reviewed all AI-assisted work. For example, LLM-generated code and figures were verified and tested for correctness by authors and all LLM-assisted proofs were checked line by line by the authors.
We take responsibility for the final content of this work, including text, claims, or artifacts produced with the aid of generative AI.


\bibliography{myIEEEabrv, bib_cumulative}
\bibliographystyle{abbrvnat}

\section*{Checklist}

\begin{enumerate}

  \item For all models and algorithms presented, check if you include:
  \begin{enumerate}
    \item A clear description of the mathematical setting, assumptions, algorithm, and/or model. [Yes. The paper clearly describes the mathematical setting, assumptions, and theoretical framework underlying the proposed results. The experimental section and associated appendices also describe the graph classes, datasets, experimental procedures, and evaluation metrics used to assess the theoretical findings.]
    \item An analysis of the properties and complexity (time, space, sample size) of any algorithm. [Not Applicable.]
    \item (Optional) Anonymized source code, with specification of all dependencies, including external libraries. [No. All experiments are reproducible based on information given in the numerical results section and associated appendices without access to the code used to produce them.]
  \end{enumerate}

  \item For any theoretical claim, check if you include:
  \begin{enumerate}
    \item Statements of the full set of assumptions of all theoretical results. [Yes.  All assumptions for theoretical results are supplied in either the main body of the paper or in the appendices.]
    \item Complete proofs of all theoretical results. [Yes. All proofs of theoretical results are supplied in either the main body of the paper or in the appendices.]
    \item Clear explanations of any assumptions. [Yes. Explanation and definition of all assumptions are supplied in either the main body of the paper or in the appendices.]
  \end{enumerate}

  \item For all figures and tables that present empirical results, check if you include:
  \begin{enumerate}
    \item The code, data, and instructions needed to reproduce the main experimental results (either in the supplemental material or as a URL). [Yes. All experiments are reproducible based on information given in the numerical results section and associated appendices.]
    \item All the training details (e.g., data splits, hyperparameters, how they were chosen). [Yes. All experimental details are provided in the numerical results section and associated appendices.]
    \item A clear definition of the specific measure or statistics and error bars (e.g., with respect to the random seed after running experiments multiple times). [Yes. All experimental specifics are included in the numerical results section and associated appendices.]
    \item A description of the computing infrastructure used. (e.g., type of GPUs, internal cluster, or cloud provider). [Yes. Description of computing infrastructure is included in numerical result appendices.]
  \end{enumerate}

  \item If you are using existing assets (e.g., code, data, models) or curating/releasing new assets, check if you include:
  \begin{enumerate}
    \item Citations of the creator If your work uses existing assets. [Yes. All datasets used in the paper are properly cited and credited.]
    \item The license information of the assets, if applicable. [Yes. All datasets used in the paper are properly cited and credited including inclusion of license information.]
    \item New assets either in the supplemental material or as a URL, if applicable. [No. The paper focuses on theoretical results with numerical illustrations.]
    \item Information about consent from data providers/curators. [Not Applicable.]
    \item Discussion of sensible content if applicable, e.g., personally identifiable information or offensive content. [Not Applicable.]
  \end{enumerate}

  \item If you used crowdsourcing or conducted research with human subjects, check if you include:
  \begin{enumerate}
    \item The full text of instructions given to participants and screenshots. [Not Applicable.]
    \item Descriptions of potential participant risks, with links to Institutional Review Board (IRB) approvals if applicable. [Not Applicable.]
    \item The estimated hourly wage paid to participants and the total amount spent on participant compensation. [Not Applicable.]
  \end{enumerate}

\end{enumerate}

\clearpage
\appendix
\thispagestyle{empty}

\onecolumn

\section{Auxiliary Definitions}
\label{extra_defs}
\textbf{Oriented incidence matrix.} For an arbitrary orientation of an undirected graph the oriented incidence matrix \( \tilde{B} \in \mathbb{R}^{n \times m} \) is defined as
\begin{equation}
  \tilde{B}_{ij} =
    \begin{cases}
    -1 & \text{if edge \(e_j\) leaves node \( v_i \),} \\
      1 & \text{if edge \(e_j\) enters node \( v_i \),} \\
      0 & \text{otherwise}.
    \end{cases}
    \label{B}
\end{equation}

\textbf{Signed adjacency matrix.} For a directed graph $G$, we may preserve the directional relationship in the line-graph $L(G)$ via the {signed adjacency matrix} $\tilde{A}_{L(G)} \in \bbR^{m \times m}$, defined as:
\begin{equation}
\tilde{A}_{ij} =
\begin{cases}
 1 & \parbox[t]{0.65\columnwidth}{if edges \(e_i\) and \(e_j\) have the same orientation to the shared vertex \(v,\)} \\[0.5ex]
-1 & \parbox[t]{0.65\columnwidth}{if edges \(e_i\) and \(e_j\) have opposite orientations to the shared vertex \(v,\)} \\[0.5ex]
 0 & \text{otherwise}.
\end{cases}
\label{Atilde}
\end{equation}

\textbf{Cut distance.} The cut distance between two graphons $\bbW_1$ and $\bbW_2$ is defined in terms of the cut norm as
\begin{equation}
\label{cutdistance}
\delta_{\square}(\bbW_1, \bbW_2)
=
\inf_{\phi}
\left\| \bbW_1^{\phi} - \bbW_2 \right\|_{\square},
\end{equation}
where the infimum is taken over all measure-preserving bijections $\phi$, and $\bbW_1^{\phi}(x,y) = \bbW_1(\phi(x), \phi(y))$ \cite{lovasz2012large}.

\textbf{Stepping Operator.} Given a kernel \textbf{K} and a partition $P=(S_1, S_2, ..., S_q)$ of $[0,1]$ into finite measurable sets, we define the {stepping operator} $\mathbf{K}_P$ by
\begin{equation}
    \mathbf{K}_P(x,y) = \frac{1}{\lambda(S_i)\lambda(S_j)}\int_{S_i \times S_j}\mathbf{K}(u,v)dudv
    \label{stepping op}
\end{equation}
for $x \in S_i$, $y \in S_j$ \citep{lovasz2012large}. We note that the stepping operator is a linear operator and contractive with respect to the cut norm, i.e.,  $||\bbK_P||_{\square} \leq ||\bbK||_{\square} $ \citep{lovasz2012large}.

\begin{proof}
Define the unitary operator $V_\phi$ on $L^2([0,1])$ by 
$(V_\phi f)(x) = f(\phi(x))$. Then for any $f \in L^2([0,1])$,
\[
(T_{\mathbf{W}^\phi} f)(x) 
= \int_0^1 \mathbf{W}(\phi(x), \phi(y)) f(y)\, dy 
= (V_{\phi^{-1}} T_{\mathbf{W}} V_\phi f)(x),
\]
so $T_{\mathbf{W}^\phi} = V_{\phi^{-1}} T_{\mathbf{W}} V_\phi$. 
Since $V_\phi$ is unitary, $T_{\mathbf{W}^\phi}$ and $T_{\mathbf{W}}$ 
are unitarily equivalent and therefore isospectral.
\end{proof}

\section{Proof of Lemma \ref{eig}}
\label{isospectral}
\begin{proof}
From the definition of the oriented incidence matrix $\tilde{B}$ \eqref{B} and signed adjacency matrix $\tilde{A}_{L(G)}$ \eqref{Atilde}, we have the following identities \citep{stanic2025eigenspaces}:
\begin{align}
    \ccalL &= \tilde{B}\tilde{B}^T, \label{eqn:lap} \\
    \tilde{A}_{L(G)} &= \tilde{B}^T\tilde{B} -2I_m, \label{eqn:line}
\end{align}
where $\tilde{B} \in \reals^{n \times m}$. 
By singular value decomposition,
\begin{align*}
    \ccalL &= U\Sigma^2U^T,\\
    \tilde{A}_{L(G)} &= V\Sigma^2V^T -2I_m.
\end{align*}
Hence, $ \tilde B \tilde B^\top$ and $\tilde B^\top \tilde B$ have the same nonzero eigenvalues $\{\sigma_i^2\}_{i=1}^r$, where $r = \mathrm{rank}(\tilde B) = n -c$ where $c$ is the number of connected components. Using identity \eqref{eqn:line},
we conclude that the eigenvalues of $\tilde A_{L(G)}$ are given by $\sigma_i^2 - 2$ for $i = 1,\ldots,r$, together with the eigenvalue $-2$ with multiplicity $m - r$. For a connected graph $G$, it holds that $r = n - 1$.


As $G$ is bipartite, there exists partition $V = V_+ \sqcup V_-$ that divides nodes on both sides of the graph. Define the diagonal matrix $D \in \mathbb{R}^{m \times m}$ with entries according to the orientation chosen in $\tilde{B}$ \eqref{B}
\[
D_{ee} = 
\begin{cases}
+1 & \text{if edge $e$ is oriented from $V_+$ to $V_-$},\\
-1 & \text{if edge $e$ is oriented from $V_-$ to $V_+$}.
\end{cases}
\]

For bipartite graphs, \(D\) reverses the signs induced by inconsistent edge orientations, yielding
\[
D\tilde A_{L(G)}D = A_{L(G)}.
\]
Hence \(\tilde A_{L(G)}\) and \(A_{L(G)}\) are diagonally similar and therefore isospectral. Thus, the eigenvalue correspondence above also holds for $A_{L(G)}$.
\end{proof}

\section{Proof of Lemma \ref{degtographon}}
\begin{proof}
Let $m_n = |E_n|$, and write the degrees of $\{G_n\}$ in decreasing order as
$d_{1,n} \ge d_{2,n} \ge \cdots$. Set
\[
p^{(n)}_i = \frac{d_{i,n}}{m_n}.
\]
By assumption, $p^{(n)} \to p$ in $\ell^2$. Following the line-graphon framework of
\citep{kandanaarachchi2024graphons}, clique structures induced by vertex degrees
play a central role in the asymptotic structure of line graphs.

For each vertex $v_{i,n}$ of $G_n$, the set of edges incident to $v_{i,n}$ forms a clique in the line-graph $L(G_n)$ of size $d_{i,n}$. Thus the line graph may be viewed as the union of vertex-induced cliques, one for each vertex of $G_n$.

Fix $K \ge 1$ and consider only the $K$ largest-degree vertices of $G_n$. These vertices induce $K$ cliques in $L(G_n)$ with edge-normalized sizes
\[
p^{(n)}_1,\ldots,p^{(n)}_K.
\]
Since $G_n$ is simple, two such vertex-induced cliques intersect in at most one line-graph vertex, corresponding to the edge joining the two original vertices if such an edge exists. Hence the total contribution of all pairwise overlaps to the induced
graphon has measure at most $O(K^2/m_n)$. Since modifying $o(m_n^2)$ adjacency entries changes the induced graphon by
$o(1)$ in $L^1$ norm, and
$\|\cdot\|_\square \le \|\cdot\|_1$, this modification is negligible in cut
norm. Therefore, after an asymptotically negligible modification in cut norm,
the contribution of the first $K$ vertices is asymptotically equivalent
to a disjoint union of $K$ cliques with edge-normalized sizes
$p^{(n)}_1,\ldots,p^{(n)}_K$, in the sense of
\citep[Lemma~8.2]{janson2016graph}. Let $U_{K,n}$ denote the $K$-block disjoint-clique graphon obtained from
this disjoint-clique approximation, with block measures given by the normalized clique sizes
$p_1^{(n)},\ldots,p_K^{(n)}$.

It remains to control the contribution of the remaining vertices. The number of line-graph edges induced by vertices outside the first $K$ is bounded above by
\[
\sum_{i>K} \binom{d_{i,n}}{2}
\leq \frac{1}{2}\sum_{i>K} d_{i,n}^2.
\]
After normalization by $m_n^2$, this contribution is bounded by
\[
\frac12\sum_{i>K}\left(\frac{d_{i,n}}{m_n}\right)^2
=
\frac12\sum_{i>K}(p_i^{(n)})^2.
\]
Since $p^{(n)} \to p$ in $\ell^2$, for every $\varepsilon>0$ we may choose $K$ large enough that
\[
\sup_{n \text{ large}} \sum_{i>K} (p^{(n)}_i)^2 < \varepsilon.
\]
Thus the line-graph structure not captured by the first $K$ cliques is negligible in cut norm.

For fixed $K$, convergence $p^{(n)}_i \to p_i$ for $1 \leq i \leq K$
implies convergence of the corresponding $K$-block disjoint-clique graphons in cut
distance. Indeed, if $U_K$ denotes the limiting $K$-block disjoint-clique graphon, then
\[
\delta_\square(U_{K,n},U_K)
\le
\|U_{K,n}-U_K\|_1
\to 0,
\]
since the graphons differ only on block regions whose measures converge to
zero.

Moreover, since $p\in\ell^2$, the omitted tail blocks contribute vanishing
cut norm as $K\to\infty$. Therefore the finite-block approximations converge
to the disjoint-clique graphon $U_p$ associated with the limiting clique
profile induced by $p$. Then,
\[
\delta_\square(W_{L(G_n)},U_p)
\le
\delta_\square(W_{L(G_n)},U_{K,n})
+
\delta_\square(U_{K,n},U_K)
+
\delta_\square(U_K,U_p).
\]

For fixed $K$, the first term tends to zero by the asymptotic
disjoint-clique approximation above, the second term tends to zero by the
finite-block convergence argument, and the third term vanishes as
$K\to\infty$ by the tail estimate, yielding
\[
\delta_\square(W_{L(G_n)}, U_p) \to 0.
\]

By the characterization of disjoint-clique graphons via mass partitions
\citep[Section~7]{janson2016graph}, the graphon $U_p$ is uniquely
determined up to measure-preserving equivalence by the limiting clique
profile induced by $p$.
\end{proof}
\label{lem4}

\section{Proof of Lemma \ref{deterministic}}
\label{determ_proof}
\begin{proof}
Define the unitary operator $V_\phi$ on $L^2([0,1])$ by 
$(V_\phi f)(x) = f(\phi(x))$. Then for any $f \in L^2([0,1])$,
\[
(T_{\mathbf{W}^\phi} f)(x) 
= \int_0^1 \mathbf{W}(\phi(x), \phi(y)) f(y)\, dy 
= (V_{\phi} T_{\mathbf{W}} V_{\phi^{-1}} f)(x),
\]
so $T_{\mathbf{W}^\phi} = V_{\phi} T_{\mathbf{W}} V_{\phi^{-1}}$. 
Since $V_\phi$ is unitary, $T_{\mathbf{W}^\phi}$ and $T_{\mathbf{W}}$ 
are unitarily equivalent and therefore isospectral.
\end{proof}

\section{Proof of Lemma \ref{lgbounds}}
\begin{proof}
Let $\{G_n\}$, $G_n = (V_n, E_n)$, be a sequence of graphs that satisfies the square-degree property \eqref{eq: sqdeg}, and let $\{L(G_n)\}$ be the associated line-graph sequence converging to a graphon $\mathbf{U}$. Let ${\mathbf{U}_{n}}$ be the graphon induced by $L(G_n)$. By Lemma \ref{deterministic}, $\lambda_i(T_{\bbU^\phi}) = \lambda_i(T_{\bbU})$ for all $\phi \in \bbPhi$. Applying Prop. \ref{weyl's} and Prop. \ref{bound}, we obtain,
    \begin{align*}
    |\lambda_i({\mathbf{U}_{n}})-\lambda_i({\mathbf{U}})| 
    &= \inf_{\phi \in \bbPhi} |\lambda_i(T_{\bbU_n}) - \lambda_i(T_{\bbU^\phi})| \\
    &\leq \inf_{\phi \in \bbPhi}\|T_{\mathbf{U}_{n}} - T_{\mathbf{U}^\phi}\|_{2} \\
    &\leq \inf_{\phi \in \bbPhi} \sqrt{8\|\mathbf{U}_{n} - \mathbf{U}^\phi\|_{\square}} \\
    &= \sqrt{8\,\delta_{\square}(\mathbf{U}_{n}, \mathbf{U})} \to 0
    \end{align*}
as $n \to \infty$, where the last inequality uses Prop.~\ref{bound} and convergence follows from $\delta_{\square}(L(G_n), \mathbf{U}) \to 0$ \citep{kandanaarachchi2024graphons}.
\end{proof}
\label{lem3.6}

\section{Proof of Thm. 
\label{thm1_proof}
\ref{laplacesparse}}
\label{thm1_proof}
\begin{proof}
Let $\{G_n\}$ be a sequence of bipartite graphs satisfying the square-degree property \eqref{eq: sqdeg} with convergent normalized degree profile \eqref{degprofile}. Let $G_{n_1}=(V_{n_1}, E_{n_1})$ and $G_{n_2}=(V_{n_2}, E_{n_2})$ be samples from this sequence with size-normalized Laplacians $\ccalL_{n_1}$ and $\ccalL_{n_2}$, respectively. 

By Lemma \ref{degtographon} $\{G_n\}$ has an associated line-graph sequence $\{L(G_n)\}$ that converges to some graphon $\mathbf{U}$. Let $L(G_{n_1}), L(G_{n_2})$ be associated with $G_{n_1}, G_{n_2}$ and have induced graphons ${\mathbf{U}_{n_1}}$ and ${\mathbf{U}_{n_2}}$ respectively. Then by Lemma \ref{eig} and Lemma \ref{lgbounds}, for sufficiently large $n$,
   \begin{align*}
   |\lambda_i(\ccalL_{{n_1}}) - \lambda_i(\ccalL_{{n_2}})|
   &=|(\lambda_i( {\mathbf{U}_{n_1}})+2/|E_{n_1}|) - (\lambda_i({\mathbf{U}_{n_2}})+2/|E_{n_2}|)|\\
   &\leq|\lambda_i({\mathbf{U}_{n_1}}) - \lambda_i({\mathbf{U}_{n_2}})| + \frac{2}{m}\\
    &\leq|\lambda_i({\mathbf{U}_{n_1}}) - \lambda_i(\mathbf{U})| + | \lambda_i({\mathbf{U}_{n_2}}) - \lambda_i(\mathbf{U})| + \frac{2}{m} \to 0
    \end{align*}
as $n_1, n_2 \to \infty$, where $m = \min\{|E_{n_1}|,|E_{n_2}|\}$.
\end{proof}



\section{Proof of Prop. \ref{SPAlinelimit}}
\label{linegraph}

\begin{proof}
The line-graph of a depth-one rooted tree is complete, since every pair of edges shares the root as a common endpoint. More generally, the line-graph of a rooted tree decomposes into complete subgraphs corresponding to internal vertices, with adjacent cliques intersecting in a single articulation vertex associated with the shared edge.

The limiting object $SPA_{\mathrm{lim}}$ is constructed by gluing rooted trees together at a common root $r$, up to finitely many extraneous nodes. By construction, the hub centered at $r$ dominates the sparse graph, implying that the corresponding complete subgraph dominates the line-graph. Therefore, as the number of nodes $n \to \infty$,
$\ccalL(SPA_{\mathrm{lim}}) \to \mathbf{W}_{SPA_{\mathrm{lim}}}$.
\end{proof}

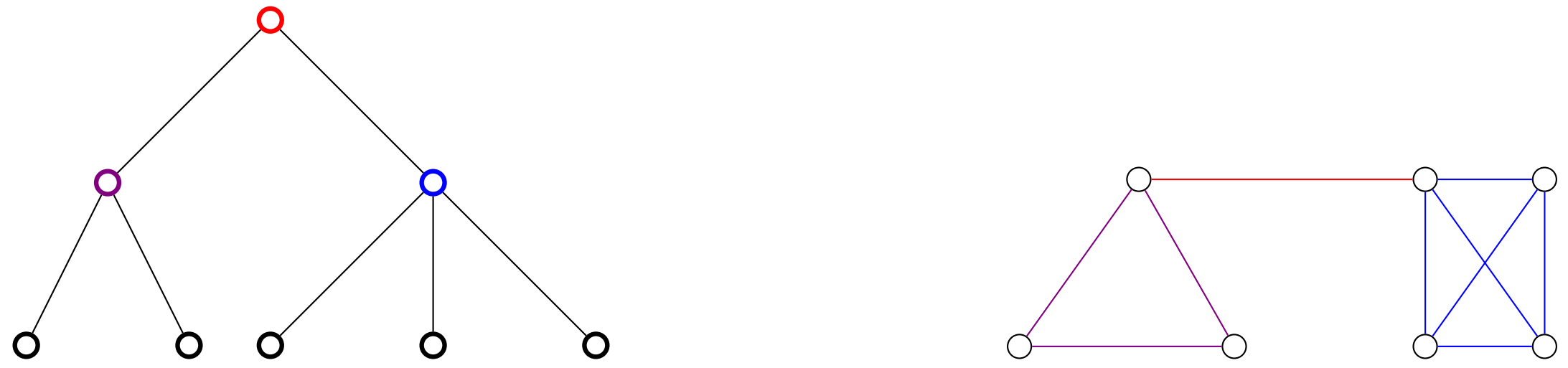
\begin{figure}[h!]
\centering
\begin{subfigure}{0.45\textwidth}
\centering
\begin{tikzpicture}[
  level distance=1.5cm,
  level 1/.style={sibling distance=3cm, color=black},
  level 2/.style={sibling distance=1.5cm, color=black},
  every node/.style={circle, draw, inner sep=1.5pt, minimum size=6pt}
]
  \node[draw=red, very thick] {}
    child {node[draw=violet, very thick] {}
      child {node[draw=black, very thick] {}}
      child {node[draw=black, very thick] {}}
    }
    child {node[draw=blue, very thick] {}
      child {node[draw=black, very thick] {}}
      child {node[draw=black, very thick] {}}
      child {node[draw=black, very thick] {}}
    };
\end{tikzpicture}
\end{subfigure}
\hspace{1cm}
\begin{subfigure}{0.45\textwidth}
\centering
\begin{tikzpicture}[
  every node/.style={circle, draw, fill=white, inner sep=2pt, minimum size=5pt},
  scale=1.1, transform shape,
  every edge/.style={draw=black, thick}
]
\node (e1) at (0,2.2){};
\node (e2) at (2.4,2.2){};
\node (e3) at (-1,0.8){};
\node (e4) at (0.8,0.8){};
\node (e5) at (2.4,0.8){};
\node (e6) at (3.4,0.8){};
\node (e7) at (3.4,2.2){};

\draw[red] (e1) -- (e2);
\draw[violet] (e1) -- (e3);
\draw[violet] (e1) -- (e4);
\draw[violet] (e3) -- (e4);
\draw[blue] (e2) -- (e5);
\draw[blue] (e2) -- (e6);
\draw[blue] (e2) -- (e7);
\draw[blue] (e5) -- (e6);
\draw[blue] (e5) -- (e7);
\draw[blue] (e6) -- (e7);
\end{tikzpicture}
\end{subfigure}
\caption{Example rooted tree (left); line-graph of rooted tree with edges colored by shared node color in original graph (right).}
\label{fig: treeline}
\end{figure}

An illustrative example of the line-graph of rooted trees is presented in Fig. \ref{fig: treeline} for clarity.

\label{linegraph}

\section{Proof of Lemma \ref{limitlap}}
\label{laplimit}

\begin{proof}
Consider the standard Hilbert space of functions on a set of vertices whose total squared magnitude is finite. We will use the set of vertices of $SPA_{lim}$, denoted $T_\infty$, and write the Hilbert space as $\ell^2(V(T_\infty))$ with inner product of elements and norm of an element defined as, \[\langle f,g\rangle = \sum_{v \in V(T_\infty)} f(v)g(v),\]  \[||f||_2 = \big(\sum_{v \in V(T_\infty)} |f(v)|^2\big)^{1/2},\] respectively.  We begin by proving the necessary properties of $\frac{1}{m}\ccalL_w^m$.

\textbf{Linear.} Let $f,g \in \ell^2(V(T_\infty))$ and $a,b$ scalars. By definition of the operator we have the following

\begin{align*}
    \frac{1}{m}\ccalL_w^m(af+bg)(y) &= \frac{1}{m}\sum_{x \sim y} A_w^m(y,x)[((af+bg)(y) - (af+bg)(x)],\\
    &= \frac{1}{m}\sum_{x \sim y} A_w^m(y,x)[a(f(y)-f(x)) + b(g(y) -g(x))],\\
    &=\frac{a}{m}\ccalL_w^mf(y) + \frac{b}{m}\ccalL_w^mg(y).
\end{align*}

\textbf{Self-adjoint.} Let $f,g \in \ell^2(V(T_\infty))$. By symmetry properties of the adjacency matrix we have,

\begin{align*}
    \langle \frac{1}{m}\ccalL_w^m f, g \rangle &= \sum_{v \in V(T_\infty)}(\frac{1}{m}\ccalL_w^m f)(v)g(v),\\
    &= \sum_{v \in V(T_\infty)}\sum_{u \sim v}\frac{1}{m}A_w^m(v,u)(f(v)-f(u))g(v),\\
    &= \frac{1}{2m}\sum_{v \in V(T_\infty)}\sum_{u \sim v}A_w^m(v,u)[(f(v)-f(u))g(v)+ (f(u)-f(v))g(u)],\\
    &= \frac{1}{2m}\sum_{v \in V(T_\infty)}\sum_{u \sim v}A_w^m(v,u)[(f(v)-f(u))(g(v) -g(u))],\\
    &= \frac{1}{2m}\sum_{v \in V(T_\infty)}\sum_{u \sim v}A_w^m(v,u)[(g(v) -g(u))(f(v)-f(u))],\\
    &= \frac{1}{m}\sum_{v \in V(T_\infty)}A_w^m(v,u)[(g(v) -g(u))f(v)],\\
    &= \langle f, \frac{1}{m}\ccalL_w^m g \rangle.
\end{align*}

\textbf{Well-defined and Bounded.} We begin by rewriting $\frac{1}{m}\ccalL_w^mf(v)$, 

\begin{align*}
    \frac{1}{m}\ccalL_w^mf(v) &= \frac{1}{m}\sum_{x \sim y} A_w^m(y,x)(f(y)-f(x)),\\
    &= \frac{1}{m}[\sum_{x \sim y} A_w^m(y,x)]f(y)- \frac{1}{m}[\sum_{x \sim y} A_w^m(y,x)]f(x),\\
    &=\frac{1}{m}d_w(y)f(y) - \frac{1}{m}A_w^mf(y).
\end{align*}

Note by construction \begin{equation}
    d_w(y) \leq 
    \begin{cases}
        A, y \neq r,\\
        \sum_kw_k, y=r,
    \end{cases}
\end{equation} 
where $r$ denotes the root node and $A$ is the max node degree for non-root nodes. Let $d_{max} = \max\{A, \sum_kw_k\}$. It is finite by construction. 

We proceed by bounding each term individually. For the first term \[\sum_y|d_w(y)f(y)|^2 \leq d_{max}^2\sum_y|f(y)|^2 < \infty\]
as $d_{max} < \infty$ by construction and $f \in \ell^2$. Thus, the first term is well-defined and bounded.

Utilizing Cauchy-Schwarz for the second term, we have \begin{align*}
\sum_y|A_w^mf(y)|^2 &= \sum_y|\sum_xA_w^m(y,x)f(x)|^2,\\
&\leq \sum_y(\sum_xA_w^m(y,x))(\sum_xA_w^m(y,x)|f(x)|^2),\\
&= \sum_yd_w(y)(\sum_xA_w^m(y,x)|f(x)|^2),\\
&\leq d_{max}(\sum_y\sum_xA_w^m(y,x)|f(x)|^2,\\
&=d_{max}\sum_x|f(x)|^2\sum_yA_w^m(y,x),\\
&=dmax\sum_x|f(x)|^2d_w(x),\\
&\leq d_{max}^2\sum_x|f(x)|^2.
\end{align*}
As $d_{max} < \infty$ and $f\in \ell^2$, this term is well-defined and bounded. 

Thus, we have a linear combination of well-defined and bounded components and $\frac{\ccalL_w^m}{m}$ is well-defined and bounded.

\textbf{Convergent over Dense Subset.} Let $f \in \ell^2(V(T_\infty))$ be finitely supported. Then, $f(v) = 0$ outside of support$(f)=\{v_1,\dots,v_n\}$. Consider the matrix interpretation \[(\frac{1}{m}\ccalL_w^mf)(v) = \sum_{u \in V(T_\infty)} \frac{1}{m}\ccalL_w^m(v,u)f(u) = \sum_{i=1}^N \frac{1}{m}\ccalL_w^m(v,v_i)f(v_i)\]
by finite support.

As $w_k^m \to w_k$ for all $k$, $A_w^m(x,y) \to A_w(x,y)$ pointwise. So, for any fixed $v$, $(\frac{1}{m}\ccalL_w^mf)(v) \to (\ccalL_\infty f)(v)$ yielding coordinatewise convergence. 

Since $f$ is finitely supported and the corresponding columns converge in $\ell^2(V(T_\infty))$ by the pointwise convergence and summability of the weights, we obtain

\[||\frac{1}{m}\ccalL_w^m f - \ccalL_\infty f||_2 \to 0.\]
Thus, we have strong convergence on finite-supported $f$ on $V(T_\infty)$. As finite support vectors in $V(T_\infty)$ are dense in $\ell^2(V(T_\infty))$, we have proved the desired property.

As $(\frac{1}{m}\ccalL_w^mf)(v) \to (\ccalL_\infty f)(v)$ for any $v$ on a dense subset of $\ell^2(V(T_\infty))$ and $\frac{1}{m}\ccalL_w^m$ is uniformly bounded, $\frac{1}{m}\ccalL_w^m$ converges strongly to $\ccalL_\infty$ \citep{eidelman2004functional}.

By strong operator convergence properties $\ccalL_\infty$ is a well-defined, boundd, self-adjoint operator on $\ell^2(V(T_\infty))$.



\end{proof}
\label{laplimit}

\section{Proof of Lemma \ref{bulkspectra}}
\label{bullkproof}

\begin{proof}
Let $T_m = \frac{1}{m}\ccalL^m_w$ and suppose $T_m$ has eigenvalues $\lambda_1^{(m)}, \dots, \lambda_n^{(m)}$. We may write the empirical spectral measure as \[\mu_m = \frac{1}{n}\sum_{i=1}^n \delta_{\lambda_i^{(m)}}.\]    By defintion, we have the following $k^{th}$ moment
\[M_k(\mu_m) = \frac{1}{n}\sum_{i=1}^n (\lambda_i^{(m)})^k.\]

We may substitute with the shifted eigenvalues of the asociated line-graph adjacency  matrix $A_m$. In accordance with Lemma \ref{eig} we write
\[M_k(\mu_m) = \frac{1}{n}\sum_{i=1}^n (\lambda_j^{(m)} + \frac{2}{m})^k.\]

By binomial expansion we have
\[M_k(\mu_m) = \frac{1}{n}tr(A_m^k) = \ccalO(\frac{k}{m}).\]

Utilizing the normalized homomorphism density we have \[tr(A_m^k) = t(C_k, H_m) \to C\]
where $H_m$ is the non-normalized line-graph adjacency matrix, $t(C_k, H_m)$ is the density of homomorphisms of $k-$cycles onto the line graph, and $C$ is a finite constant. 

Then, 
\[\frac{1}{n}tr(A_m^k) \to 0\] and 
\[\ccalO(\frac{k}{m}) \to 0\] as $k$ is a constant. Thus, $M_k(\mu_m) \to 0$. 

As all eigenvalues are in $[0,2]$, $\mu_m \to \delta_0$ and the bulk spectrum is trivial with nonzero eigenvalues as outliers.
\end{proof}
\label{bulkproof}

\section{Proof of Lemma \ref{limiteig}}
 \label{limeigproof}
\begin{proof}
Define the normalized Laplacian $\ccalL_m \;=\; \frac{1}{m}\ccalL_w^m$.
Let $A_{L(G_m)}$ be the line-graph adjacency.
Normalize by $m$ and define $A_m \;=\; \frac{1}{m} A_{L(G_m)}$.

For each fixed $i\ge 0$ (interpreting $\lambda_0(\ccalL_m)=0$),
\begin{equation}
\lambda_i(\ccalL_m) \;=\; \lambda_i(A_m) \;-\;\frac{2}{m}.
\label{eq:spa_shift}
\end{equation}
In particular, $|\lambda_i(\ccalL_m)-\lambda_i(A_m)|\le \frac{2}{m}\to 0$.

Let $\mathbf W_{SPA_{\lim}}$ be the graphon limit of the line-graph sequence
(Prop.~\ref{SPAlinelimit}), and let $\mathbf W_m$ denote the graphon induced by the adjacency $A_{L(G_m)}$ (equivalently, by the normalized line graph).
Let $T_{\mathbf W_m}$ be its associated integral operator.

By convergence of the line graphs to $\mathbf W_{SPA_{\lim}}$,
Weyl's inequality for compact self-adjoint operators yields
\[
\big|\lambda_i({\mathbf W_m})-\lambda_i({\mathbf W_{SPA_{\lim}}})\big|
\;\le\;\|T_{\mathbf W_m}-T_{\mathbf W_{SPA_{\lim}}}\|_2 \;\xrightarrow[m\to\infty]{}\; 0.
\]
Moreover, by construction $\lambda_i(T_{\mathbf W_m})=\lambda_i(A_m)$. Therefore,
\[
\lambda_i(A_m)\xrightarrow[m\to\infty]{}\lambda_i\!\left(T_{\mathbf W_{SPA_{\lim}}}\right).
\]

Combining the previous step with \eqref{eq:spa_shift} gives
\[
\lim_{m\to\infty}\lambda_i(\ccalL_m)
=\lim_{m\to\infty}\Big(\lambda_i(A_m)-\frac{2}{m}\Big)
=\lambda_i\!\left(T_{\mathbf W_{SPA_{\lim}}}\right),
\]
which concludes the proof.
\end{proof}

\label{limeigproof}

\section{Experiment Settings}
 All experiments were performed on a local machine equipped with an Apple M2 chip and 8GB of unified memory. All code was implemented in Python and run using the PyCharm IDE. The licenses and terms of use for each dataset used can be found in the following table. 

 \begin{table}[h]
\centering
\caption{Licenses and terms of use for real-world datasets used in experiments.}
\label{tab:dataset_licenses}
\begin{tabular}{p{4cm} p{8.5cm}}
\toprule
\textbf{Dataset} & \textbf{License / Terms of Use} \\
\midrule

MovieLens-1M \cite{harper2015movielens} 
& Distributed by GroupLens Research for non-commercial research use under the MovieLens dataset usage terms.{\url{https://grouplens.org/datasets/movielens/}} \\

Goodbooks-10k \cite{goodbooks10k} 
& Creative Commons Attribution-ShareAlike 4.0 International (CC BY-SA 4.0).{\url{https://github.com/zygmuntz/goodbooks-10k}} \\

Amazon Fine Food Reviews \cite{mcauley2013hidden} 
& Released through the SNAP repository for research use; distributed under CC0/Public Domain terms.{\url{https://www.kaggle.com/datasets/snap/amazon-fine-food-reviews}} \\

Book-Crossing \cite{ziegler2005improving} 
& Publicly released for research use under the original dataset distribution terms.{\url{https://www2.informatik.uni-freiburg.de/~cziegler/BX/}} \\

\bottomrule
\end{tabular}
\end{table}

\textbf{Eigenvalue Convergence on Synthetic Networks (10min Approx. Experiment Time).}

Graph sizes were
\[
\{200,300,500,800,1000,1200,1500,1800,2000,2300,2500,2800,3000,3500,4000\}.
\]
For each graph family and each graph size, we generated three independent realizations. SPA graphs were constructed by adding one edge per new node, with parameters $\alpha  =3.5$ and $\alpha= 5$.

For each sampled graph $G_n$, we computed the size-normalized Laplacian $\mathcal{L}_n = L(G_n)/|E_n|$ and extracted the three largest eigenvalues in nonincreasing order.

Errors were computed relative to the largest sampled graph in the sequence ($n=4000$) as
\[
\frac{1}{3}\sum_{i=1}^{3}
\left|
\lambda_i(\mathcal{L}_n)
-
\lambda_i(\mathcal{L}_{\mathrm{ref}})
\right|,
\]
where $\mathcal{L}_{\mathrm{ref}}$ denotes the size-normalized Laplacian of the largest graph. Reported values correspond to the mean error across runs, with shaded regions indicating one standard deviation.

\textbf{Eigenvalue Convergence on Real Data (20min Approx. Experiment Time).}

For each dataset, connected subgraphs were sampled with sizes
\[
\{500,1000,1500,2000,2500,3000,3500,4000\}.
\]
For each subgraph size and dataset, three independent realizations were generated using randomized connected subgraph sampling initialized from a randomly selected seed node. If the original graph was disconnected, sampling was restricted to the largest connected component.

For each sampled subgraph $G_n$, we computed the size-normalized Laplacian and extracted the three largest eigenvalues in nonincreasing order. Errors were computed relative to the largest sampled subgraph in the sequence ($n=4000$) as
\[
\max_{1 \leq i \leq 3}
\left|
{\lambda_i(G_n)}
-
{\lambda_i(G_{\mathrm{ref}})}
\right|,
\]
where $G_{\mathrm{ref}}$ denotes the sampled subgraph with $4000$ nodes.

\textbf{GNN Stability Across Graph Sizes (1min Approx. Experiment Time).}

All experiments used bipartite graphs with $5000$ nodes in each partition. Subgraph sizes were
\[
\{500,1000,2000,3000,4000\},
\]
where the size parameter denotes the number of sampled edges. For each graph family and subgraph size, $300$ independent edge-induced subgraphs were generated by uniformly sampling edges from the full graph and retaining their incident nodes.

The graph families considered were SPA with $\alpha \in \{2,4,5\}$ and a balanced double-star bipartite graph. The balanced double-star graph consists of one hub node in each partition, each connected to all nodes in the opposite partition.

Node features were constant, $x_i = 1$ for all nodes. Scalar regression targets were generated synthetically by sampling
\[
y_i \sim \mathcal{N}(0,1),
\]
followed by shifting the labels of the two bipartite partitions by $\pm 2$. Training supervision was provided on $50\%$ of the nodes, with the remaining nodes used for evaluation.

The GNN architecture consisted of a linear layer with hidden dimension $32$, followed by a linear output layer. Models were trained for $40$ epochs using the Adam optimizer with learning rate $0.01$.

For each graph family, a single model was trained on the full graph. The trained model was then evaluated without retraining on edge-induced subgraphs of varying size. Relative RMSE discrepancy was computed as
\[
\frac{
\left|
\mathrm{RMSE}_{\mathrm{sub}} -
\mathrm{RMSE}_{\mathrm{full}}
\right|
}{
\mathrm{RMSE}_{\mathrm{full}}
},
\]
where $\mathrm{RMSE}_{\mathrm{full}}$ denotes the test error of the model evaluated on the full graph and $\mathrm{RMSE}_{\mathrm{sub}}$ denotes the test error of the same trained model evaluated on an edge-induced subgraph.

Reported values correspond to the median relative RMSE discrepancy across the sampled subgraphs, with shaded regions indicating the interquartile range. A random seed $R=51$ was fixed across NumPy, Python, and PyTorch.

\label{compute}

\section{Additional Experiments}
\label{addexp}

All experiments were performed on a local machine equipped with an Apple M2 chip and 8GB of unified memory. All code was implemented in Python and run using the PyCharm IDE.

\subsection{Spectral Correspondence Between Laplacian and Line-Graph Adjacency (10sec Approx. Experiment Time)}
\label{eigalign}
We evaluate the spectral relationship between Laplacian eigenvalues of bipartite graphs and adjacency eigenvalues of their line graphs on three synthetic graph families: bipartite Erd\H{o}s--R\'enyi, bipartite stochastic block models (SBM), and bipartite Barab\'asi-Albert (BA) graphs. All graphs are generated with balanced partitions of size $n_1 = n_2 = 40$.

Edges in the Erd\H{o}s--R\'enyi model are sampled across the bipartition with probability $p = 0.08$, while the SBM uses cross-block probability $0.15$ and zero within-block connectivity. The BA model uses attachment parameter $m = 3$ on the bipartition.

\begin{figure}[ht!]
\centering
        \includegraphics[width=0.39\textwidth]{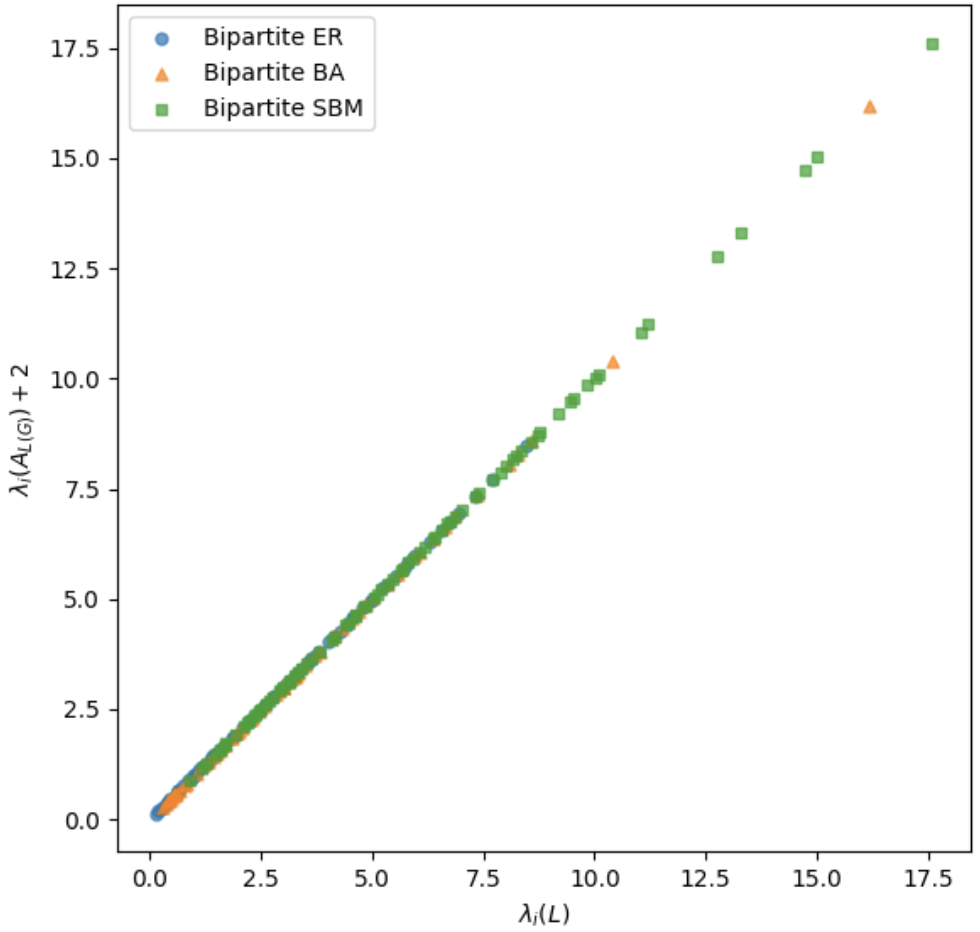} \includegraphics[width=0.6\textwidth]{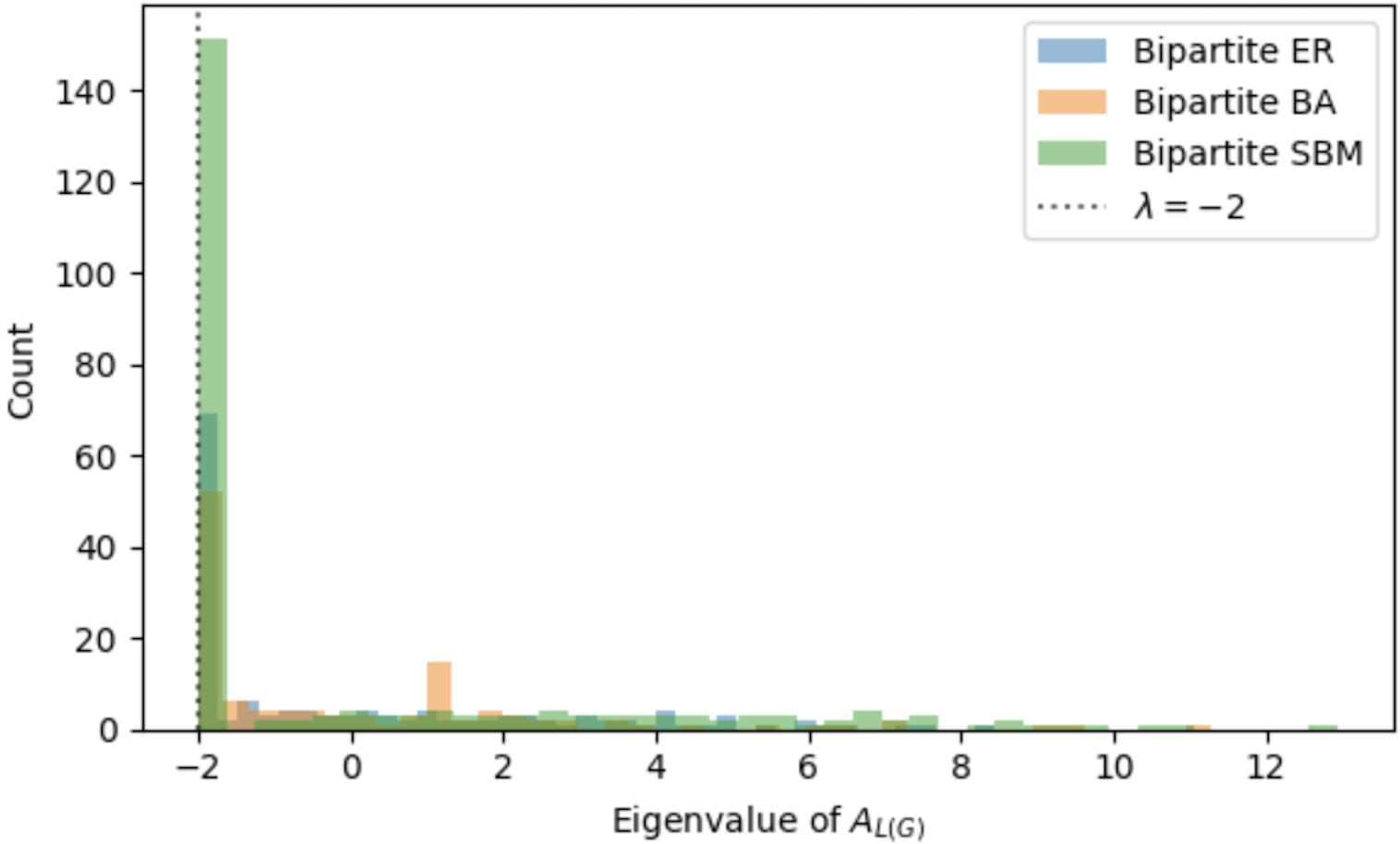}
        \caption{(Left) Scatter plot comparing Laplacian eigenvalues $\lambda_i(\mathcal{L})$ of bipartite graphs with shifted adjacency eigenvalues $\lambda_i(A_{L(G)}) + 2$ of their line graphs, illustrating the spectral correspondence predicted by Lemma~\ref{eig}. (Right) Histogram of adjacency eigenvalues of the line graph, highlighting the high-multiplicity eigenvalue at $\lambda = -2$.}
\label{fig: spectra1}
\end{figure}

For each graph, we compare Laplacian eigenvalues (excluding the trivial zero eigenvalue) with shifted line-graph adjacency eigenvalues $\lambda_i(A_{L(G)}) + 2$. The multiplicity structure of the line-graph spectrum at $\lambda = -2$ is also analyzed, consistent with Lemma~\ref{eig}. We compute spectra via full eigendecomposition of the relevant matrices using standard linear algebra routines. When necessary, we extract the largest connected component of each graph prior to analysis. Eigenvalues are sorted in ascending order, and line graphs are constructed using the standard incidence-based definition. All results are reported for individual graph instances unless otherwise specified, as the spectral identities under consideration are deterministic for fixed inputs. Results are visualized in Figure~\ref{fig: spectra1}.

\label{eigalign}

\subsection{GCN Stability Across Graph Families (1min Approx. Experiment Time)}
As an additional empirical investigation, we consider the same transferability
experiment using a standard graph convolutional network (GCN). We consider the
following graph families: SPA with $\alpha \in \{2,4,5\}$ and a balanced
double-star bipartite graph. For each family, a GCN is trained once on the full
graph using fixed random node features and supervision on $50\%$ of the nodes.
The trained model is then evaluated without retraining on edge-induced
subgraphs obtained by randomly sampling edges and retaining their incident
nodes.

For each sampled subgraph, the same trained GCN is evaluated using the graph
induced by the sampled edges. We measure stability using the relative
discrepancy between the prediction error on the sampled subgraph and that on
the full graph,
\[
\frac{
\left|
\operatorname{RMSE}_{\mathrm{sub}}
-
\operatorname{RMSE}_{\mathrm{full}}
\right|
}{
\operatorname{RMSE}_{\mathrm{full}}
}.
\]
The experiment is repeated over multiple independent edge samples at each
sample size, and the median and interquartile range of the resulting
discrepancies are reported.

The GCN uses the standard degree-normalized adjacency-based propagation
operator rather than the size-normalized Laplacian considered in the main
theoretical analysis. Consequently, this experiment is not intended as a
direct empirical validation of Theorem~\ref{laplacesparse}. Rather, it provides
a complementary comparison of learned-representation stability under
edge-induced graph restriction using a conventional GNN architecture.

\begin{figure}[t]
    \centering
    \includegraphics[width=0.8\linewidth]{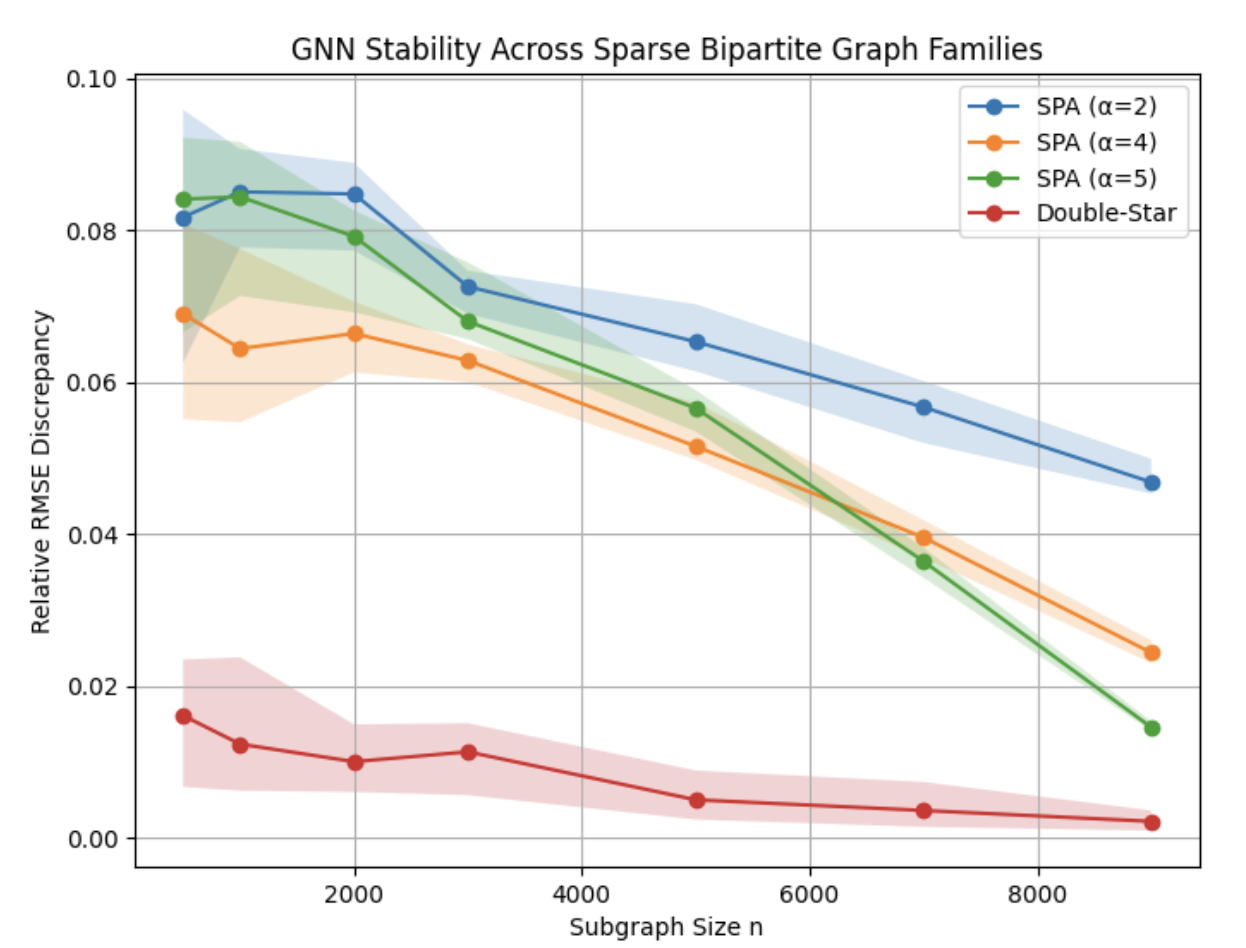}
    \caption{
    Transferability of a standard GCN under edge-induced sampling.
    The model is trained on the full graph and evaluated without retraining on
    sampled subgraphs. Curves show the median relative RMSE discrepancy, with
    shaded regions denoting the interquartile range, over repeated edge
    samples.
    }
    \label{fig:gcn-transferability}
\end{figure}

Figure~\ref{fig:gcn-transferability} reports the resulting relative RMSE
discrepancies as a function of the number of sampled edges. The qualitative
decrease in discrepancy with increasing sample size provides additional
evidence that increasing the amount of edge-induced information can improve
transferability of learned representations across graph sizes.

\label{gcn_exp}

\subsection{Square-degree Property of Real-world Data (10min Approx. Experiment Time)}
We evaluate the square-degree property (Sq) on the real-world datasets used in previous experiments: MovieLens-1M, Goodreads-10k, Amazon Fine Food, and Book-Crossing. For each dataset, we compute the square-degree ratio

$$
c_n = \frac{\sum_{i=1}^{n} d_i^2}{\left(\sum_{i=1}^{n}d_i\right)^2},
$$

where \(d_i\) denotes the degree of vertex \(v_i\).

To investigate how the ratio varies with graph size, we extract connected subgraphs containing \(n\in\{500,1000,1500,2000,2500,3000,3500,4000\}\) vertices from each dataset. At each size, we generate three samples using different random seeds and compute the square-degree ratio for each sampled graph. We report the mean ratio across the three samples, with shaded regions indicating one standard deviation.

For comparison, we consider two standard graph families with known square-degree behavior: cycles and superlinear preferential attachment (SPA) trees. For a cycle on \(n\) vertices, every vertex has degree two, giving \(c_n=1/n\), which converges to zero as \(n\to\infty\). In contrast, for the standard superlinear SPA tree model with attachment exponent \(\alpha>1\), the maximum-degree vertex asymptotically captures the total degree mass, yielding \(c_n\to 1/4\) almost surely. We include \(c_n=1/n\) as the cycle reference and \(c_n=1/4\) as the asymptotic SPA reference. 

\begin{figure}[t]
    \centering
    \includegraphics[width=0.8\linewidth]{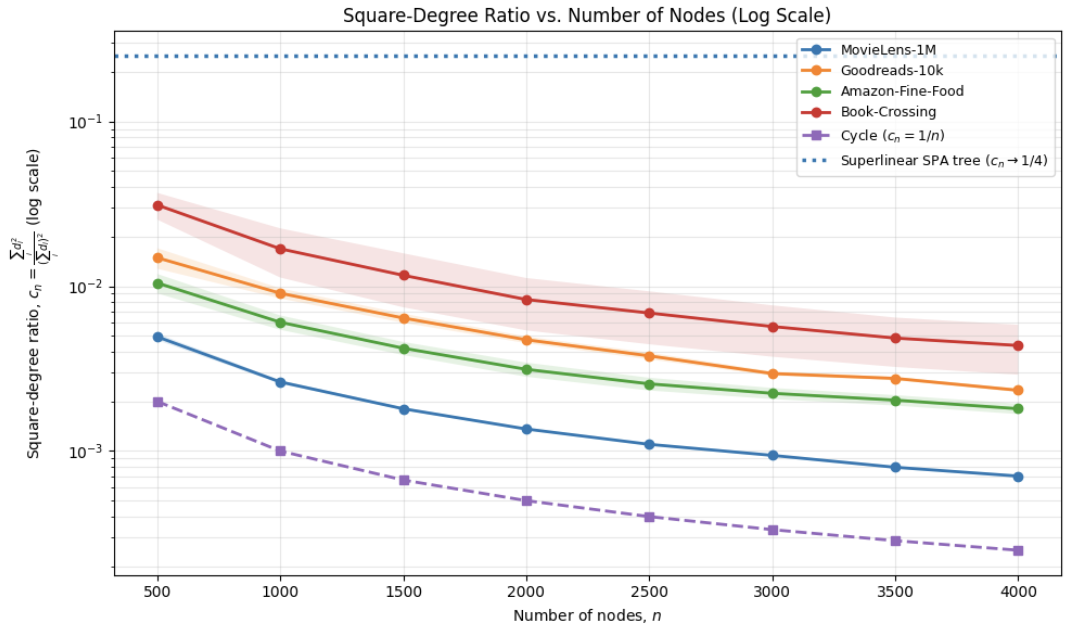}
   \caption{Square-degree ratio \(c_n=\sum_i d_i^2/(\sum_i d_i)^2\) versus the number of vertices \(n\) for four real-world bipartite graph datasets, compared with the cycle reference \(c_n=1/n\) and the asymptotic reference \(c_n\to 1/4\) for superlinear SPA trees.}
    \label{fig:realsq}
\end{figure}

This experiment illustrates the finite-size behavior of the square-degree ratio across real-world datasets and standard graph families. Since the square-degree property is an asymptotic condition, these finite-size comparisons alone do not establish whether a real-world graph sequence satisfies \ref{eq: sqdeg}. Rather they provide empirical evidence that the sampled graphs exhibit the square-degree behavior required by Thm. ~\ref{laplacesparse} at the graph sizes considered in our experiments. Results can be seen in Fig. \ref{fig:realsq}.

\label{realsq}

\subsection{Result Sensitivity to Exact Bipartiteness (10min Approx. Experiment Time)}
In a finite-size sensitivity experiment, we investigated how departures from exact bipartiteness affect the normalized-Laplacian spectrum of star graphs. We considered star graphs with
\[
n \in \{200,300,500,800,1000,1200,1500,1800,2000,
2300,2500,2800,3000,3500,4000\}
\]
vertices. Each unperturbed star is connected and bipartite, with $n-1$ edges. To introduce controlled violations of bipartiteness, we added edges between pairs of leaves, leaving the vertex set and all original edges unchanged. We considered three positive perturbation levels, corresponding to adding $\lfloor \rho(n-1)\rfloor$ edges for $\rho\in\{0.01,0.05,0.10\}$, representing $1\%$, $5\%$, and $10\%$ of the original edge count, respectively. The unperturbed case, $\rho=0$, served as the reference.

For each graph size, we computed the top three normalized-Laplacian eigenvalues of the unperturbed star and compared them with the corresponding eigenvalues of each perturbed graph. At each positive perturbation level, we generated three randomized perturbations by sampling distinct leaf pairs, using different random seeds across runs. We then calculated the mean absolute eigenvalue difference
\[
E(n,\rho)
=
\frac{1}{3R}
\sum_{r=1}^{R}\sum_{k=1}^{3}
\left|
\lambda_k\!\left(\mathcal{L}_{n,\rho}^{(r)}\right)
-
\lambda_k\!\left(\mathcal{L}_{n,0}\right)
\right|,
\qquad R=3,
\]
where $\mathcal{L}_{n,\rho}^{(r)}$ denotes the normalized Laplacian of the $r$-th perturbed graph, $\mathcal{L}_{n,0}$ denotes the normalized Laplacian of the unperturbed star, and the eigenvalues are ordered according to the eigenvalue-selection procedure used in our implementation. We report the mean of this error across the three runs for each graph size and perturbation level, with shaded bands representing one standard deviation across runs.

\begin{figure}[t]
    \centering
    \includegraphics[width=0.8\linewidth]{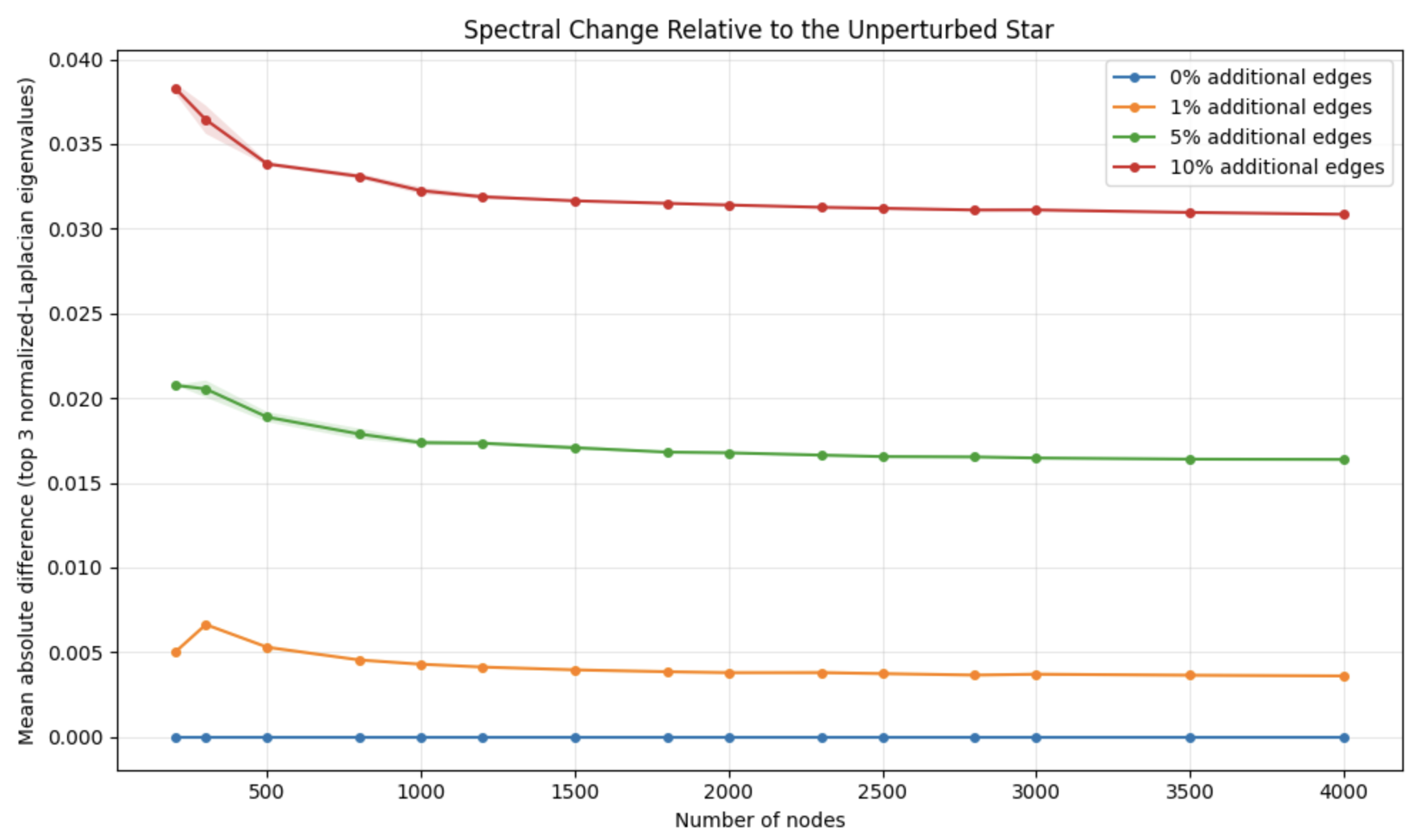}
   \caption{}
    \label{fig:perturb}
\end{figure}

The results in Fig. \ref{fig:perturb} show a clear relationship between perturbation level and spectral deviation. Across the tested graph sizes, the $5\%$ and $10\%$ perturbations produce larger mean absolute eigenvalue differences than the $1\%$ perturbation.  Although the differences decrease as the graph size increases, they remain visibly nonzero over the tested range, with the curves appearing to level off for larger graphs. These observations suggest that even relatively small departures from exact bipartiteness can alter the normalized-Laplacian eigenvalues of interest, and that the magnitude of this spectral change depends on the extent of the perturbation.

This experiment provides empirical evidence of spectral sensitivity to structural perturbations, but it does not establish that spectral convergence fails for non-bipartite graph sequences. In particular, convergence to a different limiting spectrum remains possible. 

\label{perturb_bipartite}

\end{document}